\documentclass[
 amsmath,amssymb,
 aps,
aps,
onecolumn
]{revtex4-2}
\usepackage[utf8]{inputenc}
\usepackage{amsmath}
\usepackage{graphicx}
\usepackage{appendix}
\usepackage{color}
\usepackage{hyperref} 
\usepackage{comment}
\usepackage{tikz}
\usetikzlibrary{arrows.meta}

\begin{document}
\title{Top Rank Statistics for Brownian Reshuffling}
\author{Zdzislaw Burda}\email{zdzislaw.burda@agh.edu.pl} 
\affiliation{AGH University of Krakow, Faculty of Physics and Applied Computer Science, \\
al. Mickiewicza 30, 30-059 Krak\'ow, Poland}

\author{Mario Kieburg}\email{m.kieburg@unimelb.edu.au}
\affiliation{
University of Melbourne,
School of Mathematics and Statistics,
813 Swanston Street,
Parkville, Melbourne VIC 3010,
Australia;\\
Institut Mittag-Leffler, Djursholm, SE-182 60, Sweden}

\begin{abstract}
We study the dynamical aspects of the top rank statistics of particles performing Brownian motions on a half-line, 
which are ranked by their distance from the origin. For this purpose, we introduce an observable $\Omega(t)$
which we call the overlap ratio. The average overlap ratio is equal to the probability that a particle that is on the top-$n$ list at some time will also be on the top-$n$ list after time $t$.
The overlap ratio is a local observable that is concentrated at the top of the ranking and does not require
the full ranking of all particles. In practice, the overlap ratio is easy to measure. We derive an analytical formula for the average overlap ratio for a system of $N$ particles in the stationary state that undergo independent Brownian motion on the positive real half-axis with a reflecting wall at the origin and a drift
towards the wall. In particular, we show that for $N\rightarrow \infty$, the overlap ratio takes a rather simple form
$\langle \Omega(t)\rangle = {\rm erfc}(a \sqrt{t})$ for $n\gg 1$ with some scaling parameter $a>0$.
This result is a very good approximation even for moderate sizes of the top-$n$ list such as $n=10$.
Moreover, we observe in numerical studies that the overlap ratio exhibits universal behavior in many dynamical systems including geometric Brownian motion, Brownian motion with asymptotically linear drift, 
the Bouchaud-Mézard wealth distribution model, and Kesten processes. We conjecture that the universality
 holds for a broad class of one-dimensional stochastic processes.

\end{abstract}

\maketitle

\section{Introduction}

Rankings are important tools for data analysis; see~\cite{Ranking}. Almost everything can be ranked, and often is: the richest people, the largest companies, the largest cities, the best universities, the income, the revenue, etc. The rankings are usually concentrated on the extreme values of a data set, such as the top or bottom, for instance, those who are leaders. An important and natural question is how such a ranking will evolve when the data set underlies some stochastic process. Such dynamical aspects of ranking statistics
have recently attracted attention~\cite{BGFSBBB,BKMS,IPGB,WoKu,DCLMN,KM,LD,MS2024,DHJX}.

Rankings appear very naturally in physics and other disciplines. When considering a gas of distinguishable particles, for instance, one can ask which of these particles is the closest or the furthest away from a specific point and how that evolves in time.  Certainly, when particles on a line strongly repel
each other, the ranking will always remain the same. However, once one goes to two and higher dimensions, this is no longer the case. Even when the particles have interactions like the Coulomb interaction, it is known that the distance of the particles furthest away from a point (say, the center of mass of all particles) behaves independently, see~\cite{DDMS2017}, so that overtaking in the ranking is permissible. Extreme occupations in the fluid phase of zero range processes~\cite{EM2008} and extreme heights in higher dimensional random landscapes~\cite{GKT2007} follow Gumbel statistics~\cite{Gumbel}, which suggest that the collection of these extreme events behaves effectively independently and opens the question of how their rankings develop over time. Another prominent set of models describes quantities, say $w(t)$, with dynamics driven by the rule of proportioned growth~\cite{G}, for example, wealth, in terms of multiplicative stochastic processes. In such a setting, the logarithm $x(t)= \ln w(t)$ evolves according to an additive stochastic process, where the instantaneous growth rates $\Delta x(t) =x(t+\Delta t)-x(t)$ at any time $t$ are independent of or weakly dependent on $x(t)$. In this case, the process has a stationary state in which the growth rate $x$ obeys an exponential law and, as a consequence,
the quantity $w=e^x$ follows a power law (Pareto distribution)~\cite{LS1,LS2}. Again, the question of the dynamics of the ranking of the top $n$ and/or the bottom $m$ is very natural.

\begin{figure}
\centering
\includegraphics[width=0.6\textwidth]{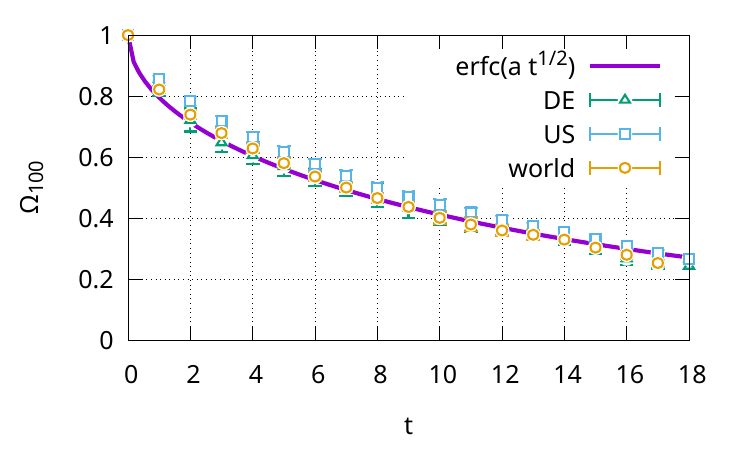} 
    \caption{Data points represent empirical data presented in \cite{BKMS}. Time $t$ is measured in years. The solid line represents theoretical curve for a model of Brownian reshuffling derived in the present work (\ref{omega.t}).
    \label{fig:empirical}}
\end{figure}

The motivation for conducting the research presented in this article comes from an observation made in \cite{BKMS} on the overlap ratio of the ranking lists of the richest people in the world.
In figure \ref{fig:empirical} we show the data points representing the overlap ratio of the $100$-top lists for lists separated by $t$ years,
which were calculated in~\cite{BKMS}. The overlap
ratio is the percentage of people who appeared simultaneously on the list in a given year and on the list after $t$ years, which is then averaged over different initial years; see also Fig.~\ref{fig:tizpicture} for a schematic particle picture. It was observed in~\cite{BKMS} that after proper rescaling, the data collapses onto a single curve suggesting universal behavior. The question is therefore whether the shape of the empirical data can be understood theoretically. This has been the goal in the present paper, where we calculate
the overlap ratio analytically in a simplified model based on a random walk, which we believe captures very well the reshuffles in the leader rankings, not only in the richest rankings but also in other rankings. The theoretical curve is shown as a solid line in figure \ref{fig:empirical}. It is given by the function ${\rm erfc}(a \sqrt{t})$ with a single free parameter, see Eq.~\eqref{omega.t}.   

The problem of ranking the top or bottom participants belongs to the field of extreme value statistics, which has recently found much interest in statistical physics and mathematics; see the review~\cite{MPS2020}. When it comes to ranking dynamics, one needs to choose a metric. The most popular metrics are Spearman $\rho$-distance~\cite{Spearman} and Kendall $\tau$-distance~\cite{Kendall}. Most of these measures take into account the total ranking of all particles. Hence, they are numerically time consuming when the set of particles is very large, and they are mostly insensitive to changes in the extreme rankings. This is the reason why we will consider the observable of the overlap ratio introduced in~\cite{BKMS}. It is a particular form of the overlap coefficient in set theory~\cite{set-distances} (also known as Szymkiewicz–Simpson coefficient). 
The overlap ratio is the percentage overlap coefficient of the top-$n$ lists at times $t_1$ and $t_2=t_1+t$.
A probabilistic interpretation of the average overlap is the probability that an element of the top-$n$ list at time $t_1$ will remain in the top-$n$ list at time $t_2=t_1+t$. There are also other set-theoretical distances; see~\cite{set-distances} and references therein, but most are more involved and less intuitive. Therefore, we stick with the simplest choice of the overlap ratio.

There are several reasons why the overlap ratio is advantageous. Firstly, it is a local quantity and does not require a ranking of the whole list. One can therefore concentrate separately on the top (leaders) or bottom (losers) of the distribution or on the bulk (typical members). Second, it is easily measurable. Finally, we discovered that it reflects universal behaviors. Demonstrating this last, and in our opinion, the most important advantage, will be the main goal of this paper. The idea is that universality should show once the system is in a stationary state and is large enough. Despite the global stationarity of the state, there will be some local fluctuations in the ranking. These fluctuations, we claim, are universal. We would like to underline that the goal of the present work is not to prove universality but to derive an analytical form for the overlap ratio, which would explain the empirical observation in Fig.~\ref{fig:empirical}. We corroborate our claim then with numerical simulations of various stochastic processes in one dimension.

To compute the universal benchmark, we propose to analytically study a rather simple toy model. We consider $N$ particles on the positive real half-axis that undergo independent Brownian motions. To find a stationary state, we introduce a reflective wall at the origin and add a negative constant drift to this wall. This interplay of barrier (wall), drift, and diffusion prevents the particles furthest from the origin from continuing to rise forever, which would forbid a reshuffling that would otherwise be induced by the stochastic nature of the dynamics. We compute the overlap ratio for this stochastic process of 
the top-$n$ particles at finite $N$ and then take the limit $N\to\infty$. The results we obtain will then be compared with the overlap ratio of the top $n$ lists measured for various stochastic processes that we numerically simulate. 
The dynamic properties of order statistics under diffusion conditions were also discussed in \cite{LD,MS2024}.

The article is structured as follows. In Sect.~\ref{sec:definitions}, we formulate the problem and sketch the main
ideas. In Sect.~\ref{sec:diffusion.model}, we review the model of a Brownian motion of $N$ particles on the positive real line with a reflective wall at the origin and a drift towards this wall. The evolution of ranks in the
stationary state of this model will serve as a prototype for Brownian reshuffling. 
Although the stationary state does not show apparent time dependence, the ranking undergoes a non-trivial evolution. In Sect.~\ref{sec:brownian}, we develop an analytic method to calculate the overlap ratio for diffusion in one dimension. In Sect.~\ref{sec:steady}, we apply this method to derive the overlap ratio for the prototype model,
that is, for diffusion on the half-axis with a constant drift. We perform an asymptotic analysis of the expression for the overlap ratio for an infinitely large number of particles. Therein, we derive expressions for the top-$n$ ranking with $n$ fixed, as well as in the limit $n\to\infty$. These results are compared with discrete-time Brownian motion in Sect.~\ref{sec:num}. To highlight the universality of the results, we also compare them with the ranking statistics of other stochastic processes, such as Brownian motion with a position-dependent drift and a lower soft barrier, geometric (multiplicative) Brownian motion, the Bouchaud-Mézard wealth distribution model~\cite{BM}, and Kesten processes~\cite{K,B}; see Sect.~\ref{sec:univ}. We also simulate diffusion
in a quadratic potential (Ornstein–Uhlenbeck process) and in a logarithmic potential, which exhibit different 
behaviors of the overlap ratio. In Sect.~\ref{sec:conclusions}, we summarize our findings and point out open problems that have arisen from our investigations. Some analytical considerations for Brownian motion on the positive real line, the overlap ratio, and details of some asymptotic analysis calculations can be found in the Appendices~\ref{app:HK_derivation},~\ref{app:sum_rules}, and~\ref{app:omega_large_n}, respectively.

\section{Basic definitions and formulation of the
problem \label{sec:definitions}}

We consider a system of $N$ elements characterized by a real quantity that can be used to order them
from largest to smallest. We might think, for example, of the wealth of people, the number of inhabitants of cities, the size, or value of firms, etc. Denote this quantity at time $t$ by $x(t)$. For the
whole system, we have $\{x_1(t),\ldots, x_N(t)\}$.
Upon sorting $x_{\sigma_t(1)}(t) \ge x_{\sigma_t(2)}(t) \ge \ldots \ge {x}_{\sigma_t(N)}(t)$, 
we can determine the ranking of the
elements at time $t$: $i \rightarrow \sigma_t(i)$, 
where $\sigma_t$ is a permutation of indices that assigns the rank to the element $i$ at time $t$. 
The element corresponding to rank $j$ can be found by inverse permutation $\sigma_t^{-1}(j)$.
As the system evolves over time, the ranking changes. To capture the rate of changes, we define
the probability $P(j;t_1|k;t_2)$ that the element that was in position $j$ in the ranking at time $t_1$ will be in position $k$ in the ranking at time $t_2$. The probability $P(j;t_1|k;t_2)$ will be called the probability of reshuffling.

\begin{figure}
\begin{tikzpicture}
\draw [thick,black](-1.5,0) -- (6.5,0);
\begin{scope}[every node/.style={circle,fill,inner sep=0.5pt}]
\node (1) at (0.2,0) {1};
\node (2) at (1.5,0) {2};
\node (3) at (2,0) {3};
\node (4) at (3.2,0) {4};
\node (5) at (3.8,0) {5};
\node (6) at (5.7,0) {6};
\end{scope}
\begin{scope}[every node/.style={circle,thick,draw,inner sep=0.5pt,red}]
\node (a) at (-0.5,0) {\phantom{1}};
\node (b) at (0.7,0) {\phantom{1}};
\node (c) at (2.8,0) {\phantom{1}};
\node (d) at (2.4,0) {\phantom{1}};
\node (e) at (4.9,0) {\phantom{1}};
\node (f) at (4.2,0) {\phantom{1}};
\end{scope}
\begin{scope}[>={Stealth[red]},
					every node/.style={fill=white,circle},
					every edge/.style={draw=red,very thick}]
\path[->] (1) edge[bend right=60] (a);
\path[->] (2) edge[bend right=60] (b);
\path[->] (3) edge[bend right=60] (c);
\path[->] (4) edge[bend right=60] (d);
\path[->] (5) edge[bend right=60] (e);
\path[->] (6) edge[bend right=60] (f);
\end{scope}
\end{tikzpicture}

\caption{Schematic picture of six particles that move along stochastic process on a line and are allowed to interchange their order. The initial position at time $t=0$ is indicated by black filled bullets while their position after the time $t$ is given by red open circles. The arrows highlight which particle moves onto what position. The overlap would be in this example $\Omega_1(0,t)=0$, $\Omega_2(0,t)=1$, $\Omega_3(0,t)=2/3$, and $\Omega_4(0,t)=\Omega_5(0,t)=\Omega_6(0,t)=1$.}
    \label{fig:tizpicture}
\end{figure}

For the ranking at time $t$, we can define the top-$n$ list as a list of elements such that
$L_n(t) = (\sigma^{-1}_t(i))_{i\le n}$.  
It is an ordered list of the $n$ leaders of the ranking. Similarly, we can also define a set of elements that are on the top-$n$ list as $T_n(t) = \{\sigma^{-1}_t(i)\}_{i\le n}$, which includes only information about elements that are present in the list but not about their position in the list. The pace of leader reshuffling can be measured by the overlap ratio \cite{BKMS},
\begin{equation}\label{overlap_def}
\Omega_n(t_1,t_2) = \frac{|T_n(t_1)\cap T_n(t_2)|}{n}
= \frac{1}{n} \sum_{j=1}^n \Theta\left(n - \left(\sigma_{t_2} \circ \sigma^{-1}_{t_1}\right)(j)\right),
\end{equation}
which is a particular example of the overlap coefficient used in set theory~\cite{set-distances}.
$\Theta(x)$ is the Heaviside step function, which means that it vanishes for $x < 0$ and is unity otherwise. We have illustrated in Fig.~\ref{fig:tizpicture} how the overlap ratio is related to the positions of the particles and their behavior over time.

On the one hand, the overlap coefficient is relatively easy to measure; on the other hand, it captures important information about the dynamics of leaders occupying one of the top positions. Similarly, we can define the overlap ratio
for losers, who are at the other end of the ranking
$\sigma^{-1}_t(i) \ge N-n$, or for
typical elements that are in the middle of the ranking:
$\sigma^{-1}_t(i) \in (m,m+n]$, where $m = \lfloor \mu N\rfloor-n/2$, 
for example, for $\mu=1/2$. 

The average overlap ratio is related to the 
probability of reshuffling $P(j;t_1|k;t_2)$:
\begin{equation} \label{average_overlap}
\langle \Omega_n(t_1,t_2) \rangle = 
\sum_{j=1}^n \sum_{k=1}^n P(j;t_1|k;t_2) .
\end{equation}
The sum on the right-hand side includes
only transitions between $j,k \le n$ that 
correspond to transitions between one of 
the $n$ leaders at $t_1$ and one of the $n$ leaders at $t_2$. The meaning of the average overlap is
the probability that an element present
in the top-$n$ list at time $t_1$ will be present
in the top-$n$ list at time $t_2$. In particular,
$\langle \Omega_1(t_1,t_2) \rangle$ is the
probability that the leader at time $t_1$
will be the leader at time $t_2$. 

In what follows, we are mainly interested
in systems that are in a stationary state.
As a prototype example, we consider the
diffusion of a gas of $N$ particles in
one dimension in a suitable potential $V(x)$.
The particles are ordered with respect to their
positions $x_i$ on the real axis. 
For a stationary system, the probability of reshuffling becomes a function of the time difference 
\begin{equation} \label{pij}
P(j;t_1|k;t_2) = P(j;k;t_2-t_1)
\end{equation}
and so does the average overlap ratio
\begin{equation}
\langle \Omega_n(t_1,t_2) \rangle = 
\langle \Omega_n(t_2-t_1) \rangle.
\end{equation}

We are mainly interested in the diffusion in the presence of
an asymptotically constant negative drift for large
for $x\rightarrow +\infty$ 
\begin{equation}
  \mu(x) = - V'(x) \rightarrow {\rm const.} < 0 .
\end{equation}
The reason we are interested in such a diffusion is as follows. In this case, 
the system has a stationary state with a probability distribution that
has an exponential tail for $x\rightarrow +\infty$.
\begin{equation} \label{expa}
  p_{\rm stat}(x) dx \sim e^{-\alpha x} dx .
\end{equation}
The exponent $\alpha$ depends on the asymptotic value of the drift and the diffusion constant. 
As we mentioned in the introduction, we want to use this model as a
model of fluctuating growth rates for some quantity,
$w=e^x$, which is subject to random multiplicative changes described by Gibrat's rule of
proportionate growth. Changing variables in (\ref{expa}) from $x$ to $w$ gives the Pareto law
\begin{equation}
  \tilde{p}_{\rm stat}(w) dw \sim w^{-\alpha-1} dw
\end{equation}
which is ubiquitous in economic systems and many other real world systems.
Since the map between the growth rate $x$ and the quantity $w=e^x$ is monotonous, the ranks of $w_i$'s 
follow the same dynamics as the ranks of $x_i$'s.

Since we are mainly interested in
the leaders of the ranking in this work, the details of the potential $V(x)$ are less important to us, as long as the potential warrants
the existence of a stationary state with a probability distribution having an exponential tail. 
Therefore, as a prototype model, we consider diffusion on the positive real half-axis with a constant
negative drift and a reflective wall at the origin
of the real axis, which ensures the existence of a stationary state. 
We believe that this is the simplest model of this type. The model is analytically solvable and thus allows us to establish a benchmark for a broader class of models.

\section{Diffusion model \label{sec:diffusion.model}}

Consider a gas of $N$ independent particles,
performing independent Brownian motion on $\mathbb{R}_+$ in the presence of a reflective
hard wall at the origin $x=0$ and a drift $\mu$ towards the wall.
The positions of these particles at time $t$ will be denoted by $x_j(t)$ and $j=1,\ldots,N$. 
Diffusion of such a system, with a diffusion constant $\sigma^2/2$, is described by a Fokker–Planck 
equation for the probability density $p(x,t)$ for finding particles at the position $x\geq 0$ at time $t\geq 0$,
\begin{equation}\label{PDE_tau}
	\partial_t p(x,t)=-\mu \partial_x p(x,t)+\frac{\sigma^2}{2} \partial_x^2 p(x,t) .
\end{equation}
The reflection at the wall at $x=0$ is enforced by the boundary condition 
\begin{equation}\label{boundary_tau}
	 -\mu p(0,t)+ \frac{\sigma^2}2 \partial_x p(0,t)=0,
\end{equation}
which means that there is no 
probability current passing through the wall. The initial probability 
density at time $t=0$ is
\begin{equation}\label{initial_tau}
	 p(x,0)=p_0(x)\qquad {\rm for}\quad x\ge 0.
\end{equation}
Since all particles forming the gas are subject to the same dynamics, the equations are independent
of the particle index $j$.
 
The differential equation~\eqref{PDE_tau} with boundary condition~\eqref{boundary_tau} and initial condition~\eqref{initial_tau} has been solved in~ \cite{H,AW}. It exhibits the following behavior. If $\mu>0$, the particles flow away from the origin. Thus, for a long time $t>0$, the initial condition is washed out and the hard wall becomes ineffective. The whole process becomes an ordinary Brownian motion with a drift, {\em i.e.}, 
\begin{equation}\label{mu>0.t>>0}
p_{\mu>0}(x,t\gg1)\approx\frac{1}{\sqrt{2\pi\sigma^2 t}}\exp\left[-\frac{1}{2\sigma^2 t}(x-\mu t)^2\right].
\end{equation}
Also for $\mu=0$, the diffusion wins against the initial condition, although the effect of the reflective wall is now visible
\begin{equation}\label{mu=0.t>>0}
p_{\mu=0}(x,t\gg1)\approx 2 \sqrt{\frac{1}{2\pi\sigma^2 t}}\exp\left[-\frac{x^2}{2\sigma^2 t}\right].
\end{equation}
The additional factor of $2$ compared to~\eqref{mu>0.t>>0} is due to condition $x>0$, which was irrelevant in the case of $\mu>0$.
The only situation where a stationary solution exists is for $\mu<0$. In this case, the drift to the hard wall and the diffusion 
balance each other and create a unique profile
\begin{equation}\label{mu<0.t>>0}
p_{\mu<0}(x,t\gg1)\approx p_{\rm stat}(x)=\frac{2|\mu|}{\sigma^2}\exp\left[-\frac{2|\mu|}{\sigma^2}x\right].
\end{equation}
and non-trivial rank statistics appear. From here on we concentrate on this case.

Considering~\eqref{mu<0.t>>0}, it is convenient to introduce the parameter $\alpha = -2 \mu/\sigma^2$, which captures the relative strength of the diffusion
constant and the drift. In fact, the differential equation~\eqref{PDE_tau} and its boundary condition~\eqref{boundary_tau} are scale invariant under the transformation $\mu \rightarrow \lambda \mu$, $\sigma^2\rightarrow \lambda\sigma^2 $, and $t \rightarrow t/\lambda$ for any $\lambda>0$. 
Choosing $\lambda = 1/\sigma^2$ gives $\mu \rightarrow \mu/\sigma^2$, $\sigma^2\rightarrow 1$, and $t \rightarrow \sigma^2 t$.
In these new units, the above equations take the form
\begin{equation}\label{PDE}
	\partial_t p(x,t)= \frac{\alpha}{2} \partial_x p(x,t)+\frac{1}{2} \partial_x^2 p(x,t) .
\end{equation}
and
\begin{equation}\label{boundary}
	 \alpha p(0,t)+  \partial_x p(0,t)=0\quad{\rm for}\quad t> 0 \qquad{\rm and}\qquad p(x,0)=p_0(x)\quad{\rm for}\quad x\ge 0.
\end{equation}
The solution of~\eqref{PDE} with these boundary
and initial conditions can be written in terms of the heat kernel $W(y,x,t)$, which is a conditional probability density. Then,
the probability density that a particle that has been initially at $y$ will be
at $x$ after time $t$ is given by
\begin{equation} \label{green}
    p(x,t) = \int_0^\infty p(y,0) W(y,x,t) dy ,
\end{equation}
and the heat kernel has the form \cite{H,AW}
\begin{equation}\label{FK_kernel}
W(y,x,t)= \partial_x F(y,x,t)
\end{equation}
where 
\begin{equation}
F(y,x,t) = \frac{1}{2}{\rm erfc}\left[\frac{2(y-x)-\alpha t}{\sqrt{8t}}\right]-
\frac{1}{2} e^{-\alpha x}{\rm erfc}\left[\frac{2(y+x)-\alpha t}{\sqrt{8t}}\right] .  
\label{F}
\end{equation}
for any $x,y\ge 0$ and $t>0$. The heat kernel $W(y,x,t)$ is also called Green's function or propagator.
The function $F(y,x,t)$ is the probability that a particle
that was initially in position $y$ will be in the interval $[0,x]$ at time $t$, so
it is just the cumulative distribution function for the probability distribution at time $t$ conditioned on the initial position $y$.
\begin{equation} \label{FK}
F(y,x,t)= \int_0^x W(y,x',t) dx' . 
\end{equation}
Obviously, it is $F(y,0,t)=0$ and $F(y,\infty,t)=1$.

The first term in $F(y,x,t)$ corresponds to the diffusion towards the wall, 
while the second corresponds to the diffusion away from the wall.
The initial position $y$ is seen as mirrored: $-y$ in the second term.  
The relative contribution of the two terms is $e^{\alpha x}$ as it comes from the drift factors $e^{\pm (\alpha/2) x}$ toward and away from the wall.

It is straightforward to check that the kernel satisfies the following equations
\begin{equation}
\partial_tW(y,x,t) =\frac{\alpha}{2}\partial_xW(y,x,t)+\frac{1}{2}\partial_x^2W(y,x,t)
\end{equation}
with 
\begin{equation}
			\alpha W(y,0,t)+\partial_x W(y,0,t)=0 \qquad{\rm and}\qquad
			W(y,x,0)=\lim_{t\to0} W(y,x,t)=\delta(x-y),  		
\end{equation}
for all $y,x\ge 0$ and $t>0$. The latter identity has to be understood in the weak sense, with $\delta(x-y)$ being the Dirac delta function.

For completeness, we present the derivation of the heat kernel $W(y,x,t)$  
in Appendix~\ref{app:HK_derivation}. A very important feature of this probability density is that it leads to a function that 
is independent of the initial position $y$ in the limit $t\rightarrow \infty$, {\em i.e.}, for $\alpha>0$ it is
\begin{equation}
p_{\rm stat}(x) \equiv W(y,x,\infty) \equiv \lim_{t\rightarrow \infty} W(y,x,t)=  \alpha e^{-\alpha x},
\label{ss}
\end{equation}
underlining the existence and uniqueness of the stationary state.
As a consequence, independently of the initial distribution $p(y,0)=p_0(y)$, the system
will end up in the state described by the probability density $p_{\rm stat}(x)$. Clearly, $p_{\rm stat}$ is the stationary state of the evolution operator
\begin{equation}
p_{\rm stat}(x) = \int_0^\infty p_{\rm stat}(y)  W(y,x,t) dy 
\label{eigenv}
\end{equation}
for any $y,x,t\ge 0$.

\section{Brownian shuffle}\label{sec:brownian}

\subsection{Probability of reshuffling}

In this subsection, we will describe how to calculate the probability of reshuffling (\ref{pij}). 
We will provide details for diffusion with constant drift on the half-axis, but
the method is general and can be applied to diffusion on the whole real axis or on an interval.
It also works for a position-dependent drift $\mu(x)=-V'(x)$, where $V(x)$ is a potential. The only change, in the general case,
would be that instead of the specific propagator given by (\ref{FK_kernel}) and (\ref{F}), we would have to use a
propagator that is appropriate for the potential and boundary conditions.
In particular, if diffusion takes place on the whole real axis, the integration range has to be changed 
from $[0,+\infty)$ to $(-\infty,\infty)$.

What is crucial for the following method is the statistical independence of the $N$ non-interacting particles that undergo diffusion.
We assume that the initial positions of the particles are at $y_1 \ge y_2 \ge \ldots \ge y_N$. 
The particles are indexed according to their initial values, from highest to lowest. 
The corresponding conditional $N$-particle probability distribution to find particles
at position $x_1,\ldots,x_N$ after time $t$ is then
\begin{equation}
    p(y_1,\ldots,y_N; x_1,\ldots,x_N; t) = \prod_{j=1}^N W(y_j, x_j, t),  
\end{equation}
where $x_j$ is the position of the particle $j$ at time $t>0$, which was initially at $y_j$.
We would like to stress that, unlike the positions $y_1 \ge y_2 \ge \ldots \ge y_N$, the positions $x_1,\ldots,x_N$ are not ordered. Surely, the order changes over time as the particles perform independent Brownian motions. The statistics of this change will be investigated below.

The probability that the particle $j$, which initially has the $j$-th largest value (the rank $j$), 
will have the $k$-th largest value at time $t$ (\ref{pij}) can be calculated as follows
\begin{equation}\label{trans.prob.fixed.y}
\begin{split}
P(k;j;t|y_1,\ldots,y_N)=&\int_0^\infty dx W(y_j,x,t)\frac{1}{(k-1)!}\partial_z^{k-1}\prod_{l\neq j}\big(F(y_l,x,t)+ z [1-F(y_l,x,t)]\big)\biggl|_{z=0}\\
=&\int_0^\infty dx  \oint \frac{dz}{2\pi i z^{k}}  \partial_x F(y_j,x,t) 
\prod_{l\neq j}\big(z+(1-z) F(y_l,x,t)\big) .
\end{split}
\end{equation}
In the second equality, we replaced the derivative $\partial_z^{k-1}$ with a contour integral in the complex plane 
over a  small circle around the origin $|z| = \epsilon < 1$. As we shall see, this method simplifies our calculations.

The interpretation of this formula is as follows. The $(k-1)$st derivative with respect to $z$ at $z=0$ selects the $k-1$ particles whose positions are larger than $x$, which corresponds to the position of the $j$th particle at time $t$. The remaining $N-k$ particles, excluding the $j$th particle, will then have positions smaller than $x$. Hence, the coefficient at the term $z^{k-1}$ of the polynomial obtained after multiplying all the factors in the product~\eqref{trans.prob.fixed.y}
contains all combinations of products of $k-1$ probabilities $1-F(y_l,x,t)$ for $x_l$ greater than $x$ and all combinations of products of $N-k$ probabilities $F(y_l,x,t)$ for $x_l$ smaller than $x$. 

Surely, the identities 
\begin{equation} \label{identities}
    \sum_{j=1}^N P(k;j;t|y_1,\ldots,y_N) =1 \qquad{\rm and}\qquad \sum_{k=1}^N P(k;j;t|y_1,\ldots,y_N) = 1
\end{equation}
must hold true for all initial positions $y_1 \ge y_2 \ge \ldots \ge y_N$ and all times $t>0$. The first identity reflects that the $j$th particle must be at some position in the ordering at time $t$, which is certainly always true, while the second identity states that one of the $N$ particles will always be at position $k$, which is also a true statement.
The identities~\eqref{identities} are proven in Appendix~\ref{app:sum_rules}. 

\subsection{Overlap ratio for deterministic initial positions \label{sec:overlap.brownian}}

Assume that the particles are initially at
positions $y_1> y_2 > \ldots > y_N$ and are indexed by their initial rank, that is, $\sigma_0(j) = j$. 
Let $\sigma_t(j)$ be the rank at time $t$ of the particle that initially had rank $j$.
The overlap ratio (\ref{overlap_def}) is then
\begin{equation}\label{overlap.def}
\Omega_n(t)=\frac{1}{n}\sum_{j=1}^n\Theta(n-\sigma_t(j)).
\end{equation}
The average overlap ratio after time $t$,
under this initial condition, can be expressed in terms of the transition probability~\eqref{trans.prob.fixed.y}, {\em i.e.},
\begin{equation}
\langle\Omega_n(t)\rangle_{y_1,\ldots,y_N} =
\frac{1}{n}\sum_{j,k=1}^n P(k;j;t|y_1,\ldots,y_N).
\label{omegaP}
\end{equation}
We denote the ensemble average under
the deterministic initial positions by $\langle.\rangle_{y_1,\ldots,y_N}$. 
The double sum adds up the probability that any of the top $n$ particles will still have one of the top $n$ ranks after time $t$.
Surely, the ordering inside the top $n$ list may change.

\subsection{Probabilistic initial positions}

Probabilistic initial conditions cover a more general setting. Since the particles do not interact with each other, we can restrict our considerations to Poisson initial conditions, where the $N$-particle probability density is the product of the single-particle probability densities
\begin{equation}
     p_0 (r_1,\ldots,r_N) = \prod_{j=1}^N p_0(r_j) .
    \label{independent}
\end{equation}
We note that the initial positions $r_1,\ldots,r_N$ are unordered, unlike $y_1> y_2 > \ldots > y_N$. 
The transition probability for the probabilistic initial conditions can be found in a similar way as in the case of~\eqref{trans.prob.fixed.y}. 
The only difference is that we have to include the initial ordering in the integral. We implement it in the same way as the ordering 
at time $t$ \eqref{trans.prob.fixed.y}, namely by taking the $(k-1)$st derivative with respect to the second auxiliary variable
\begin{equation} \label{trans.prob.poisson}
\begin{split}
P(k;j;t|  p_0 )= N
&\int_0^\infty dx \int_0^\infty dr W(r,x,t)p_0(r)\\
&\times \frac{1}{(k-1)!(j-1)!}\partial_w^{k-1}\partial_z^{j-1}
\left(\left[\int_{0}^{r}+z\int_r^\infty\right] 
\left[\int_{0}^{x}+w\int_x^\infty\right]
W(r',x',t)p_0(r')dx'dr'\right)^{N-1}\biggl|_{w=z=0} .
\end{split}
\end{equation}
The derivatives $\partial_w^{k-1}$ and $\partial_z^{j-1}$ can be replaced anew by contour integrals in the complex plane over circles centered at the origin and with a suitably small radius $\epsilon \ll 1$. The expression in the parentheses can be rewritten as follows $w P_z(r)+ (1-w) G_z(r,x,t)$ with
\begin{equation}
\begin{split}
   P_z(r) = P_-(r) + z P_+(r) \qquad{\rm and}\qquad G_z(r,x,t) = G_-(r,x,t) + z G_+(r,x,t),
\end{split}    \label{PGs}
\end{equation}
where $P_-(r) = \int_0^r p_0(r') dr'$ and $P_+(r) = 1 - P_-(z)$ are the probabilities of finding the particle at time $t=0$ in the intervals $[0,r)$ and $[r,\infty)$, and 
 \begin{equation}
 \begin{split} 
G_-(r,x,t)= \int_{0}^{r} F(r',x,t) p_0(r') dr' = \int_{0}^{r} \int_0^x W(r',x',t) p_0(r') dx' dr', \\
G_+(r,x,t)= \int_{r}^{\infty} F(r',x,t) p_0(r') dr' = \int_{r}^{\infty} \int_0^x W(r',x',t) p_0(r') dx' dr'
\end{split}
\label{Gs}
\end{equation}
are the probabilities that a particle starts in these respective intervals and ends up in the interval $[0,x]$ at time $t$. We would like to underline that they are not conditional probabilities; otherwise, we would have to divide by $\int_0^rp_0(r') dr'$ and $\int_r^\infty p_0(r') dr'$. However, they are proportional to those.
With the help of this notation, we obtain
\begin{equation}
\begin{split}
P(k;j;t|  p_0 ) 
&  = N \int_0^\infty dx \int_0^\infty dr W(r,x,t)p_0(r) \oint \frac{dz}{2\pi i z^j} \oint \frac{dw}{2\pi i w^k} 
\big( w P_z(r) + (1-w) G_z(r,x,t) \big)^{N-1}.
\end{split}
\label{Ppi0}
\end{equation}
To find this result, we have applied $\int_{0}^{x}W(r',x',t)dx'=1-\int_{x}^{\infty}W(r',x',t)dx'$,
\begin{equation}
\left[\int_{0}^{x}+w\int_x^\infty\right]W(r',x',t)dx'=(1-w)F(r',x,t)+w,
\end{equation}
cf., Eq.~\eqref{FK}.

Similarly to equation (\ref{identities}), it is straightforward to check that the sum rules
\begin{equation}\label{identities_pi0}
\sum_{j=1}^N P(k;j;t|  p_0 ) =1 \qquad{\rm and}\qquad \sum_{k=1}^N P(k;j;t|  p_0 ) = 1
\end{equation}
are satisfied, see Appendix \ref{app:sum_rules}. 

The overlap ratio
for the initial condition $ p_0 $ can be calculated by summing over $j,k=1,\ldots, n$ in~\eqref{Ppi0}, 
analogously to~\eqref{omegaP}. These sums are two independent geometric sums; for instance, for $1/z^j$ it leads to 
\begin{equation}
    \sum_{j=1}^N\frac{1}{z^j}=\frac{1-z^{-n}}{z-1}=\frac{1}{z-1}-\frac{1}{1-z} \frac{1}{z^n}.
\end{equation}
The integral of the first term $1/(z-1)$ vanishes as the origin is no longer a pole, so only the term $-1/[(1-z) z^n]$ survives.
A similar summation can be applied to the sum over $k$ and contour integration over $w$. Hence, we arrive at the expression
\begin{equation}
\begin{split}
\langle\Omega_n(t)\rangle_{ p_0 }
&  = \frac{N}{n} \int_0^\infty dx \int_0^\infty dr W(r,x,t)p_0(r) \oint \frac{dz}{2\pi i (1-z) z^n} \oint \frac{dw}{2\pi i (1-w) w^n} 
\big( w P_z(r) + (1-w) G_z(r,x,t) \big)^{N-1}.
\end{split}
\label{OmegaGeneral}
\end{equation}
To keep the notation simple, we denote the ensemble average under the initial condition $p(x,0)=p_0(x)$ by $\langle.\rangle_{ p_0 }$.

\section{Overlap ratio for stationary state of diffusion on half-axis \label{sec:steady}}

The calculations of the overlap ratio discussed so far were general. From now on, we will focus on the model
on the half-axis defined in Section \ref{sec:diffusion.model}, which was motivated in Section \ref{sec:definitions}.
The aim is to calculate the overlap ratio for the stationary state in this model. Once the system reaches the stationary state,
the probability distribution will remain constant; however, the order of particles on the real axis will be continuously rearranged.
In the stationary state, as follows from~\eqref{eigenv}, the single particle distribution $p(r,t)$ at any
moment in time is identical to that at the beginning $p(r,t)=p_0(r)=p_{\rm stat}(r) = \alpha e^{-\alpha r}$ (\ref{ss}). 
As a consequence, the $N$-point probability distribution is constant over time:
\begin{equation}
     p (r_1,\ldots,r_N;t) = p_0 (r_1,\ldots,r_N) = \prod_{j=1}^N p_{\rm stat}(r_j).
\end{equation}
Statistically speaking, the entire system looks the same at every moment, except that the order of the particles changes.

We denote the average overlap ratio by $\langle \Omega_n(t) \rangle_{N,\alpha}$ for the stationary state. 
Clearly, it depends on the size of the top-$n$ list, the total size of the system $N$, the stationary state parameter $\alpha$, see~\eqref{ss}, and the time elapsed between the recordings of the top-$n$ lists. Using~\eqref{OmegaGeneral}
we find 
\begin{equation}\label{omega.coeff}
    \langle \Omega_n(t) \rangle_{N,\alpha} = \frac{1}{n!(n-1)!} \partial^{n-1}_z \partial^{n-1}_w  Z_{N}^{(\alpha)}(z,w,t) \bigg|_{z=w=0},
\end{equation}
where 
\begin{equation} \label{Z}
\begin{split}
     Z_{N}^{(\alpha)}(z,w,t)  =&\frac{N}{(1-z)(1-w)}\int_0^\infty dx
     \int_0^\infty dr W(r,x,t) \alpha e^{-\alpha r} \\
     & \times \left[ w\bigl(1+(z-1) e^{-\alpha r} \bigl) + (1-w) \bigl(G_-(r,x,t) + z G_+(r,x,t)\bigl)\right]^{N-1}
\end{split}
\end{equation}
is a kind of generating function.
In the expression above, we replaced $p_0(x)$ with $p_{\rm stat}(x)=\alpha e^{-\alpha x}$ and plugged in
\begin{equation}
P_z(r)=\left[\int_0^r  +z\int_r^\infty\right]dr' \alpha e^{-\alpha r'}=1+(z-1) e^{-\alpha r}.
\end{equation}
The functions $G_\pm(r,x,t)$ are incomplete Gaussian integrals once we plug the stationary state $p_0(r)=p_{\rm stat}(r)$ and~\eqref{F} into~\eqref{Gs}, which yields complementary error functions
\begin{equation}
\begin{split}
G_-(r,x,t)
&=\frac{1}{2}\biggl(2-{\rm erfc}\left[\frac{2r+2x+\alpha t}{\sqrt{8 t}}\right]-e^{-\alpha x}\left(2-{\rm erfc}\left[\frac{2r-2x+\alpha t}{\sqrt{8 t}}\right]\right)\\
&-e^{-\alpha r}\left(2-{\rm erfc}\left[\frac{2x-2r+\alpha t}{\sqrt{8 t}}\right]\right)
+e^{-\alpha(r+x)}{\rm erfc}\left[\frac{2r+2x-\alpha t}{\sqrt{8 t}}\right]\biggl),\\
G_+(r,x,t)
&=\frac{1}{2}\biggl({\rm erfc}\left[\frac{2r+2x+\alpha t}{\sqrt{8 t}}\right]-e^{-\alpha x}{\rm erfc}\left[\frac{2r-2x+\alpha t}{\sqrt{8 t}}\right]\\
&+e^{-\alpha r}\left(2-{\rm erfc}\left[\frac{2x-2r+\alpha t}{\sqrt{8 t}}\right]\right)
-e^{-\alpha(r+x)}{\rm erfc}\left[\frac{2r+2x-\alpha t}{\sqrt{8 t}}\right]\biggl).
\end{split}
\end{equation}

Equation~\eqref{omega.coeff} shows that the overlap ratios
are essentially the Taylor series coefficients of $Z_{N}^{(\alpha)}(z,w,t)$ in variables $z$ and $w$ about the point $w=z=0$ at powers $(zw)^{n-1}$.
For instance, in the simplest case of $n=1$, the overlap ratio takes the simple form
\begin{equation} \label{omega1_N}
\langle \Omega_1(t) \rangle_{N,\alpha} = 
\alpha N \int_0^\infty dx \int_0^\infty dr W(r,x,t) e^{-\alpha r} G_-^{N-1}(r,x,t).
\end{equation}
The overlap ratio $\langle \Omega_1(t) \rangle_{N,\alpha}$ is simply the probability that the current leader will be the leader after time $t$. In a recent work~\cite{MS2024}, denoted by $S_N(t_1,t_2)$ therein, it was used for the long time analysis  of the decorrelation of the position of the leader. For long times $t\gg1$, the original ranking would have been forgotten, and the average overlap converges to $\langle \Omega_1(t) \rangle_{N,\alpha}\sim 1/N$.

In a similar way, we may derive
explicit integral expressions for $\langle \Omega_2(t) \rangle_{N,\alpha}$, $\langle \Omega_3(t) \rangle_{N,\alpha}$, {\em etc.} from $Z_{N}^{(\alpha)}(z,w,t)$. The integral expressions can be used to numerically compute the overlap ratios.

\subsection{Asymptotic analysis\label{sec:asymptotic}}

Our next goal is to analyze the asymptotic of~\eqref{Z} for $N\to\infty$ and $n\ll N$.
As we consider independent particles drawn from the stationary state $p_{\rm stat}(r)=\alpha e^{-\alpha r}$, the largest particles will follow Gumbel statistics; in particular, the distribution of the position of the $j$th top particle is
\begin{equation}
\tilde{p}_{{\rm top},j}(r)= \frac{\alpha N!}{(N-j)!(j-1)!}e^{-j\alpha r}(1-e^{-\alpha r})^{N-j}\approx\frac{\alpha}{(j-1)!}  e^{-j(\alpha r-\ln N)}\exp\left[-e^{-(\alpha r-\ln N)}\right].
\end{equation}
Therefore, the highest values are located near $\ln N/\alpha$.
The dominant contribution to the integral~\eqref{Z} for $N\rightarrow \infty$ can then be obtained by going over to the new variables
$\xi$ and $\zeta$.
\begin{equation} \label{xi_zeta}
     r = \frac{\ln N}{\alpha} + \zeta  , \quad x = \frac{\ln N}{\alpha} + \xi.
\end{equation}

The functions appearing in the integral~\eqref{Z} can be expanded in $1/N$
\begin{equation}
\begin{split}
W\left(\frac{\ln N}\alpha + \zeta, \frac{\ln N}\alpha + \xi,t\right) & =
W\left(\frac{\ln N}\alpha + \zeta, \frac{\ln N}\alpha + \xi,t\right)  =
\omega(\zeta, \xi,t) + O(1/N), \\
G_-\left(\frac{\ln N}\alpha + \zeta, \frac{\ln N}\alpha + \xi,t\right) & =
1 - \frac{1}{N} g_-(\zeta, \xi,t) + O(1/N^2), \\
G_+\left(\frac{\ln N}\alpha + \zeta, \frac{\ln N}\alpha + \xi,t\right) & =
- \frac{1}{N} g_+(\zeta, \xi,t) + O(1/N^2) \\
\end{split}
\label{Womega}
\end{equation}
where 
\begin{equation}
\begin{split}, \\
g_-(\zeta,\xi,t)  & = \frac{1}{2}\biggl[e^{-\alpha \xi}\left(2-{\rm erfc}\left[\frac{2\zeta-2\xi+\alpha t}{\sqrt{8 t}}\right]\right)+e^{-\alpha \zeta}\left(2-{\rm erfc}\left[\frac{2\xi-2\zeta+\alpha t}{\sqrt{8 t}}\right]\right)\biggl],\\
g_+(\zeta,\xi,t) & = \frac{1}{2}\biggl[e^{-\alpha \xi}{\rm erfc}\left[\frac{2\zeta-2\xi+\alpha t}{\sqrt{8 t}}\right]-e^{-\alpha \zeta}\left(2-{\rm erfc}\left[\frac{2\xi-2\zeta+\alpha t}{\sqrt{8 t}}\right]\right)\biggl].
\end{split}
\end{equation}
Moreover, we exploit $e^{-\alpha r} = e^{-\alpha \zeta}/N$ and
$g_+(\zeta,\xi,t) = e^{-\alpha \xi} - g_-(\zeta,\xi,t)$
to recast~\eqref{Z} into
\begin{equation} 
\begin{split}
     Z_{N}^{(\alpha)}(z,w,t) =& \frac{\alpha}{(1-z)(1-w)}\int_{-\ln N/\alpha}^\infty d\xi
     \int_{-\ln N/\alpha}^\infty d\zeta \left(\omega(\zeta,\xi,t) e^{-\alpha \zeta} + O(1/N)\right)\\
     & \times \left[ 1 - \frac{1}{N} \bigg(      
     w(1-z) e^{-\alpha \zeta} + z(1-w) e^{-\alpha \xi} + (1-z)(1-w) g_-(\zeta,\xi,t)\bigg) +
     O(1/N^2) \right]^{N-1} .
\end{split}
\end{equation}
Taking the limit $N\rightarrow \infty$ gives 
\begin{equation} 
\begin{split}
     Z^{(\alpha)}(z,w,t) =&  \frac{\alpha}{(1-z)(1-w)}\int_{-\infty}^\infty d\xi
     \int_{-\infty}^\infty d\zeta \omega(\zeta,\xi,t) e^{-\alpha \zeta} \\
     & \times \exp \left[ -w(1-z) e^{-\alpha \zeta} - z(1-w) e^{-\alpha \xi} - (1-z)(1-w) g_-(\zeta,\xi,t)\right] . 
\end{split}
\end{equation}
Changing the integration variables to $R=(\xi+\zeta)/2$ and $s=(\xi-\zeta)/\sqrt{2t}$ and performing the integral with respect to $R$, we finally obtain
\begin{equation}
    Z^{(\alpha)}(z,w,t) = Z_*(z,w,t_*),
    \label{zstar}
\end{equation}
where 
\begin{equation}
    t_* = \frac{\alpha^2 t}{8}
    \label{tstar}
\end{equation}
and
\begin{equation} 
     \label{zstar2}
     Z_*(z,w,t_*) = 
     \frac{e^{-t_*}}{\sqrt{\pi}} \frac{1}{(1-z)^2(1-w)^2} \int_{-\infty}^\infty ds 
 \frac{e^{-s^2}}{e^{-2 s\sqrt{t_*}}\left(\frac{1}{1-z}-\frac{1}{2}{\rm erfc}\left[\sqrt{t_*}-s\right]\right)+e^{2s\sqrt{t_*}}\left(\frac{1}{1-w}-\frac{1}{2} {\rm erfc}\left[\sqrt{t_*
}+s\right]\right)}.
\end{equation}
Interestingly, the result depends on $\alpha$ only through the rescaled time variable
$t_*$. Expanding $Z_*(z,w,t_*)$ in $z$ and $w$, we find overlaps as coefficients
at the powers $(zw)^{n-1}$. In particular, for $n=1$ it is
\begin{equation}
    \langle \Omega_1(t) \rangle_{\alpha} \overset{N\gg1}{\approx} \Omega_{*1}(t_*) =
     \frac{e^{-t_*}}{\sqrt{\pi}} \int_{-\infty}^\infty ds 
 \frac{e^{-s^2}}{e^{-2 s\sqrt{t_*}}\left(1-\frac{1}{2}{\rm erfc}\left[\sqrt{t_*}-s\right]\right)+e^{2s\sqrt{t_*}}\left(1-\frac{1}{2}{\rm erfc}\left[\sqrt{t_*
}+s\right]\right)} ,
\label{omega1}
\end{equation}
and for $n=2$ we find
\begin{equation}
    \langle \Omega_2(t) \rangle_{\alpha} \overset{N\gg1}{\approx} \Omega_{*2}(t_*) =
     \frac{e^{-t_*}}{\sqrt{\pi}} \int_{-\infty}^\infty ds e^{-s^2}
 \left(\frac{1}{q^3(s,t_*)} - \frac{e^{2s\sqrt{t_*}}+e^{-2s\sqrt{t_*}}}{q^2(s,t_*)} + \frac{2}{q(s,t_*)}\right)
 \label{omega2}
\end{equation}
where
\begin{equation}
    q(s,t_*)=e^{-2 s\sqrt{t_*}}\left(1-\frac{1}{2}{\rm erfc}\left[\sqrt{t_*}-s\right]\right)+e^{2s\sqrt{t_*}}\left(1-\frac{1}{2}{\rm erfc}\left[\sqrt{t_*
}+s\right]\right)
\label{denom}
\end{equation}
 is the denominator in the fraction in~\eqref{omega1}.
 
In principle, we can find an explicit formula for any $n$ in the form of an integral over one variable, as we illustrate below. Using~\eqref{zstar} as a starting point, we have the relation
\begin{equation}
\langle \Omega_n(t) \rangle_{\alpha} \overset{N\gg1}{\approx} \Omega_{*n}(t_*) =
\frac{1}{n}\oint \frac{dz}{2\pi i z^n} \oint \frac{dw}{2\pi i w^n} Z_*(z,w,t_*)
\label{omega_star_n}
\end{equation}
where the contours are circles centered at $z=0$ and $w=0$ with a suitably small radius $\epsilon \ll 1$.
Changing the integration variables $u=1/(1-z)-1$ and $v=1/(1-w)-1$ in~\eqref{omega_star_n} with $Z_*(z,w,t_*)$ given by~\eqref{zstar2}, we obtain
\begin{equation}
 \Omega_{*n}(t_*) = \frac{e^{-t_*}}{n\sqrt{\pi}} \int_{-\infty}^{\infty} ds e^{-s^2} 
\oint \frac{dv}{2\pi i} \oint \frac{du}{2\pi i } 
 \frac{\left(1+1/v\right)^n\left(1+1/u\right)^n}{ e^{-2s \sqrt{t_*}} u +  e^{2s \sqrt{t_*}} v + q(s,t_*)} .
\end{equation}
Applying the binomial expansion formula for the powers in the numerator and the geometric series expansion for the expression in the denominator, and choosing terms of order $1/u$ and $1/v$, we get
\begin{equation}
 \Omega_{*n}(t_*) = \frac{e^{-t_*}}{n\sqrt{\pi}} \int_{-\infty}^{\infty} ds e^{-s^2}
 \sum_{j,k=1}^n (-1)^{j+k} \binom{n}{j} \binom{n}{k} \binom{j+k-2}{j-1} e^{2(k-j) s \sqrt{t_*}} 
 \left(\frac{1}{q(s,t_*)}\right)^{j+k-1} .
 \label{omega_n}
\end{equation}
For $n=2$ this gives $\Omega_{*2}(t_*)$, cf.,~\eqref{omega2}.  In general, one can find an explicit form of the integrand for any $n$ and then perform the integration numerically. It becomes tedious as $n$ increases but is doable. Therefore, we strive for an approximation when $n\gg1$, which is given in the next section.

\subsection{Asymptotic result for long top lists }

We employ an integral representation of the third binomial in~\eqref{omega_n}
\begin{equation}
    \binom{j+k-2}{j-1} = \oint \frac{dz}{2 \pi i z} \frac{(1+z)^{j+k-2}}{z^{j-1}} =
    \oint \frac{dz}{2 \pi i z} \left(1+ z^{-1}\right)^{j-1} (1+z)^{k-1} , 
\end{equation}
where the integral is over a small circle centered at $z=0$ with a suitably small radius 
$\epsilon <1 $. Then, the binomial sums over $j$ and $k$ factorize,
\begin{equation}
 \Omega_{*n}(t_*) = \frac{e^{-t_*}}{n\sqrt{\pi}} \int_{-\infty}^{\infty} ds e^{-s^2}
  \oint \frac{dz}{2 \pi i z}
\frac{ q(s,t_*)}{(1+z^{-1})(1+z)}
 \sum_{j=1}^n \binom{n}{j} \left(-(1+z^{-1})\frac{e^{-2s\sqrt{t_*}}}{q(s,t_*)}\right)^j 
 \sum_{k=1}^n \binom{n}{k} \left(-(1+z)\frac{e^{2s\sqrt{t_*}}}{q(s,t_*)}\right)^k .
\end{equation}
The sum over $j$ can be extended by including the term for 
$j=0$, for which the contour integral gives zero, so we get 
\begin{equation}
 \Omega_{*n}(t_*) = \frac{e^{-t_*}}{n\sqrt{\pi}} \int_{-\infty}^{\infty} ds e^{-s^2}
  \oint \frac{dz}{2 \pi i z}
\frac{ q(s,t_*)}{(1+z^{-1})(1+z)}
 \left(1 - (1+z^{-1})\frac{e^{-2s\sqrt{t_*}}}{q(s,t_*)}\right)^n 
\left[\left(1- (1+z)\frac{e^{2s\sqrt{t_*}}}{q(s,t_*)}\right)^n-1\right] . 
\label{doublepole}
\end{equation}
We show in Appendix (\ref{app:omega_large_n}) that for $n\rightarrow \infty$ the integral approaches 
the limit $\Omega_{*n}(t_*) \rightarrow \Omega_*(t_*)$, which takes the form
\begin{equation} \label{exp2p}
 \Omega_{*}(t_*) = \frac{e^{-t_*}}{\sqrt{\pi}} \int_{-\infty}^{\infty} ds e^{-s^2}
  \int_{-\infty}^{\infty} \frac{ dy}{2 \pi }
\frac{ q(s,t_*)}{(1+ i  y)^2}
 \exp\left[(1+ i  y)\frac{e^{-2s\sqrt{t_*}}}{q(s,t_*)}\right]
\left(1- \exp\left[-(1+ i  y)\frac{e^{2s\sqrt{t_*}}}{q(s,t_*)}\right]\right) .
\end{equation}
We can simplify this expression by changing the integration variable $\tilde{y}=(1+iy)/q(s,t_*)$ 
\begin{equation}
 \Omega_{*}(t_*) = \frac{e^{-t_*}}{\sqrt{\pi}} \int_{-\infty}^{\infty} ds e^{-s^2}
  \int_{\gamma} \frac{d\tilde{y} }{2 \pi i \tilde{y}^2}
 \exp\left[\tilde{y} e^{-2s\sqrt{t_*}}\right]
\left(1-  \exp\left[-\tilde{y} e^{2s\sqrt{t_*}}\right]\right) .
\end{equation}
where $\gamma$ is a vertical line ${\rm Re}\, \gamma =1/q(s,t_*)$ extending from $-i\infty$ to $+i\infty$.
We notice that the dependency on $q(s,t_*)$ has essentially disappeared, as we are allowed to shift the contour parallel to the positive real axis without changing the result due to Cauchy's theorem
(as long as no simple pole is crossed by the contour while it is being shifted). 
 
\begin{figure}
    \centering
    \includegraphics[width=0.6\textwidth]{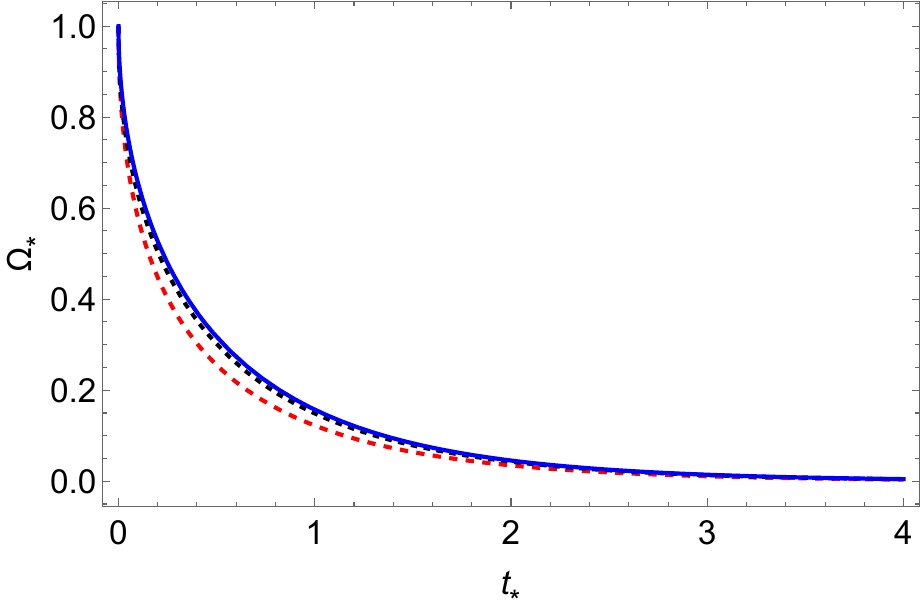} 
    \caption{Overlap ratio vs. rescaled time for $n=1,5,\infty$ (dashed, short-dashed, solid) obtained using Eqs.~\eqref{omega_n} and~\eqref{omega_star}, respectively. The rescaled time $t_*$ is given by~\eqref{tstar}. If we were to draw the corresponding curve for $n=10$, using the line width as in the plot, it would coincide with the curve for $n=\infty$.
    \label{fig:omega_star}}
\end{figure}
The integral over $\gamma$ can be calculated using the residue theorem. It is convenient to split it into two integrals,
\begin{equation}
 \Omega_{*}(t_*) = \frac{e^{-t_*}}{\sqrt{\pi}} \int_{-\infty}^{\infty} ds e^{-s^2}
 \left( \int_{\gamma} \frac{d\tilde{y} }{2 \pi i \tilde{y}^2}
 \exp\left[\tilde{y} e^{-2s\sqrt{t_*}}\right] -
\int_{\gamma} 
\frac{d\tilde{y} }{2 \pi i \tilde{y}^2}
 \exp\left[\tilde{y} (e^{-2s\sqrt{t_*}} -e^{2s\sqrt{t_*}})\right]
\right).
\end{equation}
The first integral can be calculated by closing the contour $\gamma$
to the left. In contrast, we have a case discussion for the second term, where we close the contour to the left when $s<0$ and to the right when $s>0$. The latter would yield zero, as there is no singularity on that side. Therefore, we obtain
\begin{equation}
 \Omega_{*}(t_*) = \frac{e^{-t_*}}{\sqrt{\pi}} \int_{-\infty}^{\infty} ds e^{-s^2}
\left(
e^{-2s\sqrt{t_*}} - \Theta(-s) \left(e^{-2s\sqrt{t_*}} -e^{2s\sqrt{t_*}}\right)\right)=
{\rm erfc}(\sqrt{t_*})
\label{omega_star}
\end{equation}
with $\Theta(s)$ the Heaviside step function. 

In summary, for $n\rightarrow \infty$, the mean value of the overlap ratio, $\langle \Omega_n(t) \rangle \rightarrow \langle \Omega(t) 
\rangle$, takes a very simple form
\begin{equation}\label{omega.t}
  \langle \Omega(t) \rangle = {\rm erfc}\left(\sqrt{\frac{\alpha^2 t}{8}}\right).
\end{equation}
In Fig.~\ref{fig:omega_star}, we compare the overlap ratio for finite $n$ (see Eq.~\eqref{omega_n}) to the limiting
result~\eqref{omega_star} for $n\rightarrow \infty$. Already, the curve for $n=1$ lies very close to the limiting curve for $n=\infty$, while the curve for $n=5$ touches the limiting curve almost everywhere. Thus, there is a fast convergence in $n$. The curves for
finite $n$ are obtained by the numerical integration of~\eqref{omega_n}.

\section{Numerical simulations of diffusion on half-axis}\label{sec:num}

\begin{figure}
    \centering
    \includegraphics[width=0.6\textwidth]{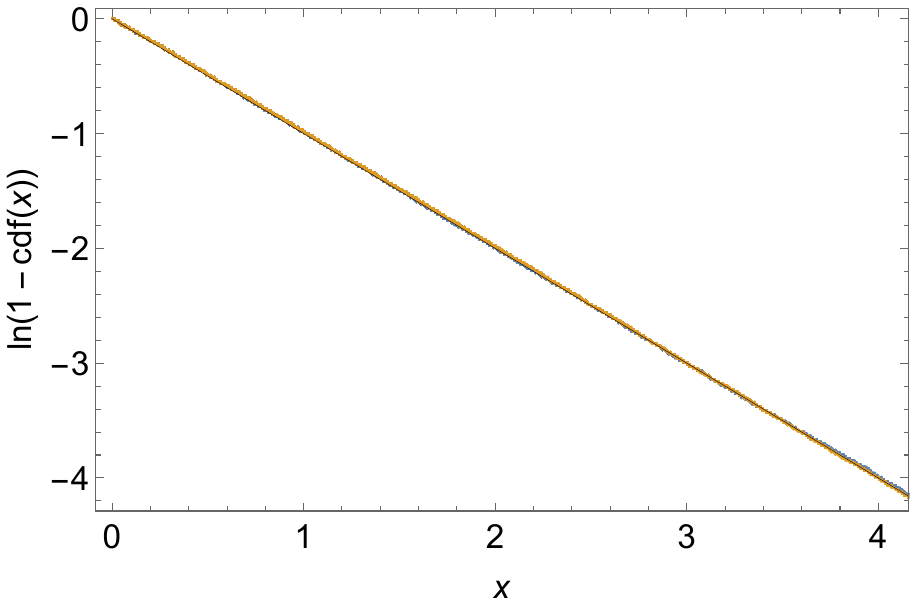} 
    \caption{The cumulative density function for the discrete
    time formalism is computed numerically from a population of $N$ particles at some
    time $k$ as a rank plot made of pairs of points $(x_i(k),i/N)$, for $i=1,\ldots,N$, where $x_i(k)$'s are sorted $x_1(k) \le \ldots \le x_N(k)$. The time $k$ is long enough to ensure that the system is in a stationary state. We plot the data for two times $k$ and $k+10^3$ to check that the system has indeed already reached a stationary state.
    On the vertical
    axis we plot $y=\ln(1-P_-(x))$ which for the exponential distribution is $y=-\alpha x$.
    In the simulation we used $N=10^5$, $\Delta t=0.05$, $\alpha=1$ and $k=10^4$. 
    Initially the particle positions were distributed according to the uniform distribution
    on $[0,1]$. The graph shows that the system has reached a stationary state, because the data points for $k$ and $k+10^3$ lie almost on top of each other. They also lie 
    on top of the curve $y=-\alpha x$ which is drawn as a thin solid line. \label{fig:cdf_num_vs_th}}
\end{figure}

In this section, we discuss Monte Carlo simulations of the same model but formulated in the discrete-time formalism, in which time is indexed by an integer. We consider a gas consisting of $N$ independent particles diffusing in one dimension, 
whose positions $x_j(k)$, $j=1,2,\ldots,N$ at time $k$ obey the following evolution equation
\begin{equation} \label{dtf}
  x_j(k) = |x_j(k-1) + g_j(k)|
\end{equation}
where for all $j=1,\ldots, N$ and $k=1,2,\ldots$ the variables
$g_j(k)$ are independently and identically normally distributed random numbers with mean 
$\mu \Delta t$ and variance $\sigma^2 \Delta t$.
If there were no absolute value on the right-hand side, the equation would describe the random walk (diffusion) on the real axis. The absolute value acts like a reflecting wall that changes the sign of $x_j(k-1)+g_j(k)$ every time it becomes negative, thus keeping the particles
on the positive half-axis. The continuum limit, where we can define the continuous physical time $t= k\Delta t$, corresponds to the double limit in which $\Delta t \rightarrow 0$ while $t=k\Delta t$ is constant. In this limit, the Fokker-Planck equation~\eqref{PDE_tau} is restored. If
we additionally choose time units so that $\sigma^2=1$, then the system will be described
by~\eqref{PDE} with $\alpha=-2\mu > 0$ that we used to study the
evolution of order statistics in the stationary state, which has an exponential
distribution, with a probability density $p_{\rm stat}(x) = \alpha e^{-\alpha x}$. For finite $\Delta t$, the stationary state in the discrete time formalism~\eqref{dtf} may slightly differ from the exponential distribution. The finite size effects are very small for $\Delta t$
of order $0.01$. This is illustrated in Fig.~\ref{fig:cdf_num_vs_th} where the cumulative distribution function computed numerically for $\Delta t=0.05$ is compared with the cumulative density function for the exponential distribution. The agreement is so good that no difference between the numerical points and the cumulative distribution for the exponential distribution can be detected with the naked eye. 

\begin{figure}
    \centering
    \includegraphics[width=0.6\textwidth]{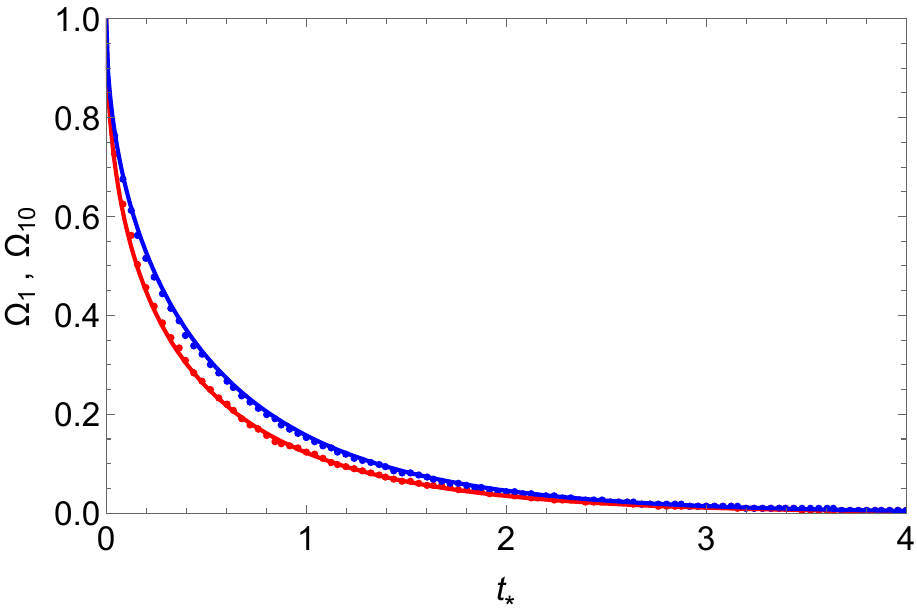} 
    \caption{Overlap ratio for $n=1$ (red dot, lower curve) and $n=10$ (blue dots, upper curve) plotted against a scaled time $t_*$, see~\eqref{tstar}, calculated in the discrete-time formalism, compared to theoretical results calculated in the continuous-time formalism using equations~\eqref{omega1} (red line) and~\eqref{omega_star} (blue line), respectively. Discrete-time data points are calculated using Monte Carlo simulations for $\alpha=1$, $N=5000$, $\Delta t_*=0.04$, repeated $10^4$ times to estimate the average 
    overlap coefficient. Error bars are smaller than the size of the dots. \label{fig:omega_th_vs_mc}}
\end{figure}

To compute the overlap ratio in the discrete-time formalism,
we simulate the evolution of a system of $N$ particles according to~\eqref{dtf}. First, we generate a stationary state and register $n$ 
leaders in this state. We treat this state as the initial state.  
We then continue the simulation and check how many leaders from the initial top-$n$ list are on the top-$n$ list at time $k$, 
{\em i.e.}, 
what the overlap between these lists is, expressed as the number of leaders present on both lists.
We repeated this experiment multiple times to calculate the average overlap and then normalize it to $n$, 
thus obtaining the average overlap ratio $\langle \Omega(k)\rangle$ over time $k$. 

The result of the Monte Carlo simulations is shown in Fig.~\ref{fig:omega_th_vs_mc} where we compare the overlap ratio for $n=1$ and $n=10$ 
with theoretical results for $n=1$ in~\eqref{omega1} and $n=\infty$ in~\eqref{omega_star} obtained within the continuous-time formalism. 
We see that the two-formalisms give consistent results. We show results for $\alpha=1$ ($n=1$ and $n=10$),
but we checked that also for other values of $\alpha$ and $n$ they are consistent.

\begin{figure}
    \centering
    \includegraphics[width=0.6\textwidth]{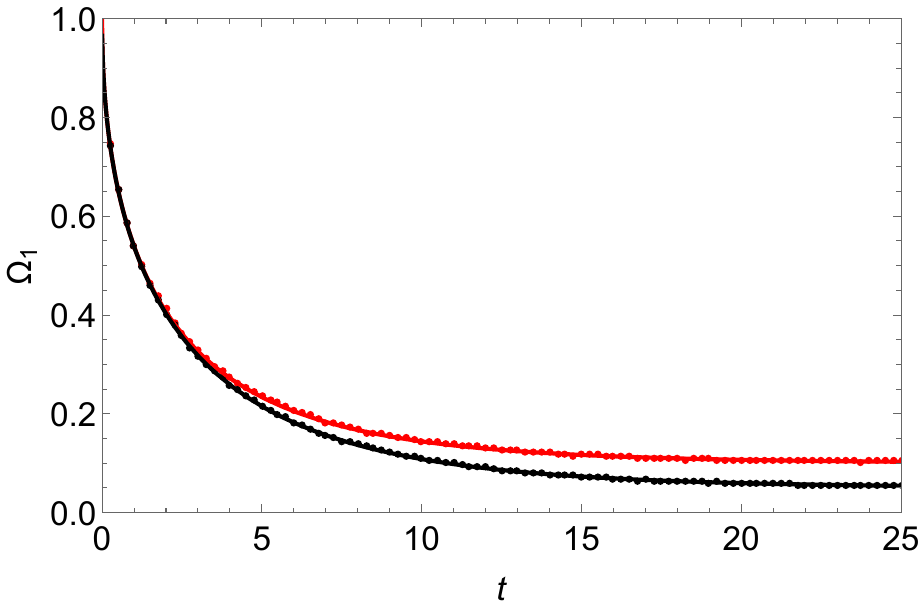}
    \caption{Overlap ratio for $n=1$ and $N=10$ (black dot, lower curve) and $N=20$ (red dots, upper curve) calculated using Monte-Carlo simulations 
    with theoretical curves obtained using continuous time formalism calculated from Eq.~\eqref{omega1_N} 
    for $N=10,20$ (solid lines). The curves asymptotically approach the limiting value $1/N$,  which is equal $0.1$ and $0.05$ for $N=10,20$, respectively. Each data point
    was calculated as an average from $5 \times 10^4$ independent Monte-Carlo simulations. 
    The size of the error bars is smaller than the size of the dots.
     \label{fig:omega_1_finite_N}    
     }
\end{figure}

The finite-$N$ corrections are also consistent in both formalisms, as shown in Fig.~\ref{fig:omega_1_finite_N}, where the average overlap ratio for $n=1$ calculated analytically in~\eqref{omega1_N} for the continuous-time formalism and numerically for the discrete-time formalism is plotted for two $N$. In general, the overlap ratio $\langle \Omega_n(t)\rangle$ should approach $n/N$ for $t\rightarrow \infty$, which is equal to the probability of randomly selecting $n$ participants from $N$. For the curves presented in the figure, the asymptotic values are $1/10$ and $1/20$, for $N=10$ and $20$, respectively.

\section{Universality of reshuffling for asymptotically constant negative drift\label{sec:univ}}

To highlight the universality of the results, we have numerically studied the overlap ratios of four one-dimensional point processes,
which find applications in various research areas. All examples effectively correspond for $x\rightarrow \infty$
to a one-dimensional diffusion in a potential $V(x)$ that asymptotically generates a negative drift
$\mu(x) = - V'(x) \sim -\alpha/2 < 0$ for $x \rightarrow \infty$ 
As we shall see, all these four cases produce quite a universal picture as far as rank dynamics is concerned, belonging to the same universality class as the model defined in Sect. \ref{sec:diffusion.model} and motivated towards the end of Sect. \ref{sec:definitions}.

For illustration and to show the limitations of the picture, 
we also provide counterexamples that lead to a different overlap behavior. 
They are diffusion in a quadratic potential, for which
the drift $\mu(x) = -V'(x) \sim -x$ increases in absolute value linearly with $x$, 
and in a logarithmic potential, for which the drift $\mu(x) = -V'(x) \sim -1/x$ tends to zero.

When comparing empirical or numerical data with our benchmark, we propose the following observation. The only essential free parameter related to the
reshuffling rate of leaders on the top-$n$ list is $\alpha$. Therefore, the practical problem when fitting data
to theoretical curves is how to identify the parameter $\alpha$. For moderately large $n$, for example $n=10$, we know that~\eqref{omega.t} is a good approximation. When integrating this quantity in time $t$ over an interval $[0,T]$, one finds for large $T\gg1$ that
\begin{equation}\label{emp.fit}
\int_0^T dt \langle\Omega(t)\rangle=\int_0^T dt\, {\rm erfc}\left(\sqrt{\frac{\alpha^2 t}{8}}\right)\sim\frac{4}{\alpha^2}.
\end{equation}
The statistical error of this quantity will be relatively low because of the large amount of data it takes; {\em i.e.}, 
it is an integral and, therefore, rather robust and weakly bin-independent like a cumulative probability density.

The first four examples below will fall into the class of processes that we claim belong to the universality class we have studied. The fifth and sixth numerical studies will not, and indeed, show deviations in the behavior of the overlap ratio.

\subsection{Example 1: Brownian motion with position-dependent drift \label{sec:example1}}

\begin{figure}
    \centering
    \includegraphics[width=0.6\textwidth]{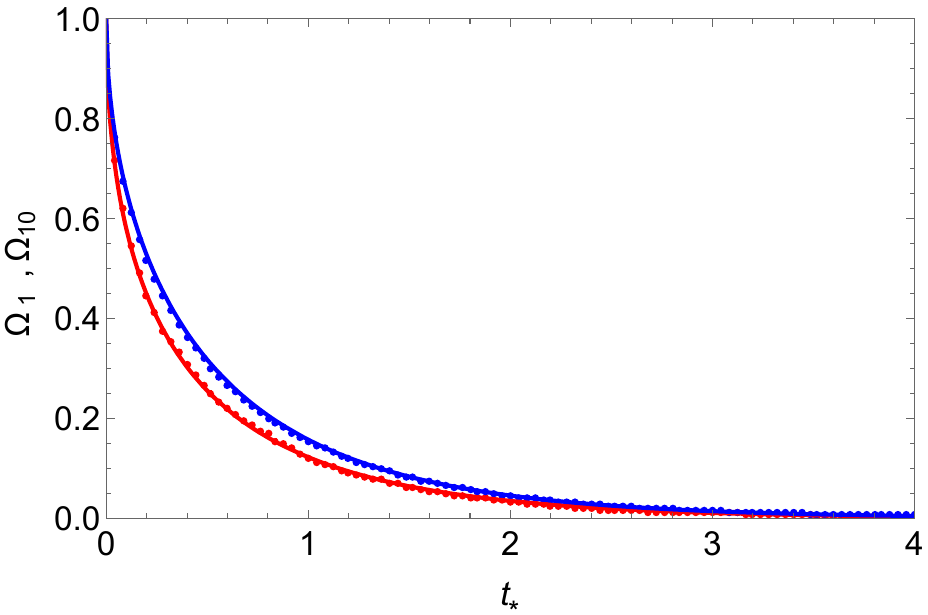}
    \caption{ Overlap ratio for $N=5000$, $\alpha=1$ for one dimensional diffusion with the drift $\mu(x)$ given (\ref{mux}) for $n=1$ (red dots, lower curve) and $n=10$ (blue dots, upper curve) obtained by averaging over $10^4$ Monte-Carlo samples, compared to the theoretical Eqs.~\eqref{omega1} (lower red line) and~\eqref{omega_star} (upper blue line).
     \label{fig:Omega_mux}    
     }
\end{figure}

As mentioned in Sect. \ref{sec:definitions} we expect the order statistics of the leaders to
be independent of the details of the potential, as long as it generates a negative 
constant drift asymptotically for
$x \rightarrow \infty$. To test this expectation, we replace the diffusion on the half axis in this section with
a diffusion in an external potential $V(x)$, which is described by the following
stochastic equation (in Itô calculus)
\begin{equation} \label{drif}
d x_j(t) = \mu(x_j) dt + dB_j(t)
\end{equation}
where $B_j(t)$, $j=1,\ldots,N$ are independent Wiener processes, and 
$\mu(x)=-V'(x) \rightarrow -\alpha/2< 0$ for $x\rightarrow \infty$. The corresponding Fokker-Planck equation
\begin{equation}
	\partial_t p(x,t)= -\partial_x \left(\mu(x) p(x,t)\right) +
    \frac{1}{2} \partial_x^2 p(x,t) 
\end{equation}
has a stationary state solution
\begin{equation}
  p_{\rm stat}(x) = \frac{1}{c_0} \exp\left[ \int_0^x 2 \mu(x') dx'\right] \qquad{\rm with}\quad c_0=\int_{-\infty}^\infty dx \exp\left[ \int_0^x 2 \mu(x') dx'\right].
\end{equation}
For large positive $x$, the probability density behaves asymptotically as $p_{\rm stat}(x) \sim e^{-\alpha x}$.
We silently assumed that stationary states exist, which imposes some conditions on $\mu(x)$, 
also for $x\rightarrow -\infty$. As an example, we consider the position dependent drift in our simulations.
\begin{equation} \label{mux}
    \mu(x) = \left\{ \begin{array}{cl} -\alpha/2, & \mbox{for}\quad x\ge 0, \\
                                       -\alpha/2+x^2/2, & \mbox{for}\quad x < 0 \end{array} \right.
\end{equation}
with $\alpha>0$. In the Monte-Carlo simulations, we used an approximated version of Eq. (\ref{drif}) 
\begin{equation}
x_j(k) = x_j(k-1) + \mu(x_j(k-1)) \Delta t + \sigma \sqrt{\Delta t} {g}_j(k).
\end{equation}
where $\hat{g}_j(k)$ are independent and identically distributed random variables generated from
a normal distribution with a zero mean and unit variance. The physical time $t$ is related to the discrete time $k$ as follows: $t=k \Delta t$. In Fig.~\ref{fig:Omega_mux}, we show, as an example, the result of Monte Carlo calculations of the overlap ratio for $n=1$ and $n=10$ in the presence of a soft barrier (\ref{mux}).
As we can see in the figure, the overlap ratio is described by
the same curve as for diffusion on $\mathbb{R}_+$. As mentioned in Sect.~\ref{sec:definitions}, for any
stationary diffusion with an asymptotically constant negative drift $-\alpha/2<0$ for $x\rightarrow \infty$, 
we expect the behavior of leaders to be shaped by an exponential tail $p_{\rm stat}(x) \sim e^{-\alpha x}$ 
and thus described by the same universal formulas for the overlap ratios found in~\eqref{omega_n} and~\eqref{omega_star}. This is exactly what can be seen in Fig.~\ref{fig:Omega_mux}.

\subsection{Example 2: Overlap ratio for multiplicative stationary processes}\label{sec:example2}

If we interpret the stochastic process $x(t)$ as a logarithm of a random variable $w(t)$, that is,
$x(t) = \ln w(t)$, then $\Delta x(t)=x(t+\Delta t)-x(t)$ is just
the growth rate of $w(t)$ for the period from $t$ to $t+\Delta t$. The quantity $w(t)$ undergoes a stochastic multiplicative process in which the growth rate is independent of $w(t)$ itself, which is known as Gibrat's rule of proportional growth~\cite{G}. 

Generally, when the total growth rate $x(t)$ for the period
from $0$ to $t$ grows unlimitedly as time $t$ increases, $x(t)$ obeys normal (non-stationary) statistics with mean and
variance increasing with $t$, and thus $w(t)$ obeys log-normal statistics.
However, when the growth of the growth rate is limited by the presence of constant negative drift (and a barrier from below), the growth rate has a stationary state with
an exponential tail, that is, $p_{\rm stat}(x) dx \sim e^{-\alpha x} dx$ for large $x$. Thus, the quantity
$w$ has a stationary state with a Pareto tail in the probability density function,
\begin{equation}
    \tilde{p}_{\rm stat}(w) dw \sim w^{-1-\alpha} dw ,
\end{equation}
which follows from changing $w=e^x$ in $p_{\rm stat}(x) dx$.

Since the map $x\rightarrow w=e^x$ is strictly monotonically increasing, the order statistics
for $w$ are identical to those of $x$. Therefore, the results~\eqref{omega_n}
and~\eqref{omega_star} still apply to a class of Pareto distributions generated as
stationary states of multiplicative stochastic processes. 

In the ensuing subsections, we provide some evidence that
the universality also extends to a broader class of mixed additive-multiplicative 
stochastic processes, where the leaders effectively undergo multiplicative
changes with an asymptotic constant drift of the effective growth rate towards smaller values.

\subsection{Example 3: Overlap ratio for the Bouchaud-M\'ezard model}\label{sec:example3}

The Bouchaud-M\'ezard model~\cite{BM} describes the distribution of wealth in a population of
individuals. The dynamics of the system is shaped by two factors: the rule of proportionate growth, which describes the growth of wealth for each individual, and the
redistribution of wealth between interacting
individuals. In the mean-field version of the model, 
the evolution of wealth is given by the following equations
\begin{equation}
    w_j(k) = (1-b) \tilde{w}_j(k) + \frac{b}{N} \sum_{l=1}^N 
    \tilde{w}_l(k),  \ j=1,\ldots,N
\end{equation}
where 
\begin{equation}
    \tilde{w}_l(k) = w_l(k-1) e^{g_l(k)}, \ l=1,\ldots,N 
\end{equation}
and
$g_l(k)$ are independently and identically distributed normal random variables with mean $\mu \Delta t$
and variance $\sigma^2 \Delta t$. The parameter $\Delta t$ is
the physical time between subsequent discrete times, which means between $k-1$ and $k$. The elements $\tilde{w}_j(k)=w_j(k-1) e^{g_j(k)}$ on the right-hand side
correspond to the multiplicative changes in wealth
described by Gibrat's rule $w_j(k-1) \rightarrow w_j(k-1) e^{g_j(k)}$. The parameter $b$ is a redistribution factor. 
In the mean-field version of the model, wealth is redistributed according to the principle of proportionality, which says that a share of each individual's wealth, proportional to wealth
itself, goes to a common pool from which all receive equal shares. The wealthiest contribute the most to the pot and receive much less from it as a result of redistribution.
This is a kind of balancing mechanism that causes a stationary state to appear in the distribution of wealth.
The parameter $b = \beta \Delta t$ depends linearly on $\Delta t$, such as the mean $\mu \Delta t$ and the variance $\sigma^2 \Delta t$. The coefficient $\beta$ is the rate of wealth redistribution. 

Actually, one is usually less interested in the absolute wealth of individuals but more in the relative one. Hence, we consider  the distribution of wealth of individual $j$ relative to the average wealth at time $k$,
\begin{equation} \label{relative}
    v_j(k) = \frac{w_j(k)}{\frac{1}{N}\sum_{l=1}^N w_l(k)} .
\end{equation}
In the continuum limit and for $N\rightarrow \infty$, the evolution equations for the relative wealth take  the form
(in the It\^o convention) 
\begin{equation} \label{dv}
    d v_j(t)= \beta(1-v_j(t)) dt + \sigma v_j(t) dB_j(t) .
\end{equation}
They are identical for all $j$, and they are  decoupled from each other for $N\rightarrow \infty$, as then $N^{-1}\sum_{l=1}^N w_l(k)$ converges to a random number that is asymptotically independent of $w_j(k)$. Therefore, one can write a Fokker-Planck equation (independent of $j$)
\begin{equation}
    \partial_t p(v,t) = \beta\partial_v\left((v-1) p(v,t)\right)  + \frac{\sigma^2}{2}
\partial^2_v \left(v^2 p(v,t)\right)
\end{equation}
The equation has a stationary state solution given by the inverse Gamma distribution with its probability density
\begin{equation} \label{inv_gamma}
    \tilde{p}_{\rm stat}(v) = \frac{c}{v^{1+\alpha}} \exp\left[-\frac{\alpha-1}{v}\right],
\end{equation}
where $\alpha = 1 + 2\beta/\sigma^2>0$ and $c=(\alpha-1)^\alpha/\Gamma(\alpha)$.
By construction, the mean of the distribution is one, $\int v p_v(v)d v = 1$, reflecting the normalization~\eqref{relative}.

 We note that the power
$\alpha=1+2\beta/\sigma^2$ is independent of the drift $\mu$. This is because $\mu$ is
identical for all $j$ and, therefore, cancels out in the definition of relative wealth~\eqref{relative}. In this model, the role of negative drift comes from wealth redistribution, controlled by $\beta$. Redistribution
has the greatest impact on reducing the relative wealth of
the richest. The dependence on $\mu$ remains in the expression for average wealth, 
which for $N\rightarrow \infty$ behaves on average like
\begin{equation}
    \left\langle \frac{1}{N} \sum_{l=1}^N w_l(t) \right\rangle = w_0 \exp \left[\left(\mu+\frac{\sigma^2}2\right)t \right] . 
\end{equation}

\begin{figure}
    \centering
    \includegraphics[width=0.6\textwidth]{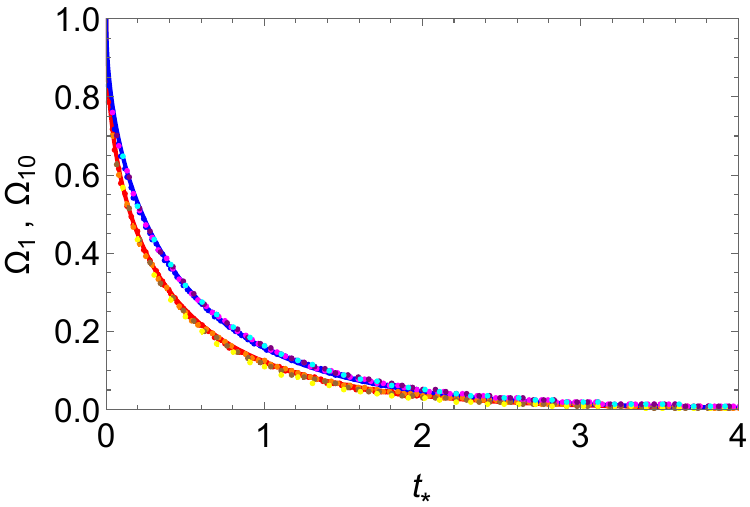}
    \caption{Overlap ratio for $N=5000$, $n=1$ (lower curve) and $n=10$ (upper curve) and $\alpha=2,3,4,5$ for the Bouchaud-M\'ezard model computed using 
    $10^4$ Monte Carlo samples. Data points are plotted against scaled time $t_* =s(\alpha) t$, where $s=0.490, 1.044, 1.731, 2.513$ for $\alpha=2,3,4,5$, respectively, as given in Eq.~(\ref{seff}). As one can see, this choice of the scaling factor $s(\alpha)$ makes 
    the points collapse to the curves given by Eqs.~\eqref{omega1} and ~\eqref{omega_star}. These curves are depicted as continuous lines that are almost invisible in the crowd of data points of different colors,
    which correspond to different $\alpha$ and $n$.
     \label{fig:BM20}    
     }
\end{figure}

For large $v$, the distribution has a Pareto tail. Considering the discussion in subsection~\ref{sec:example2}, we would, therefore, expect that our prediction should hold true in this case.
In Fig.~\ref{fig:BM20}, we show the comparison of the overlap ratio for $n=1$ and $n=10$
obtained in Monte Carlo simulations of the Bouchaud-M\'ezard model for different values of $\alpha$ and the analytical results~\eqref{omega1} and~\eqref{omega_star} for the diffusion on $\mathbb{R}_+$. To collapse the data, we rescaled the time variable $t_* = s(\alpha)t$. The scaling factors in Fig.~\ref{fig:BM20} are $s=0.48,1.03,1.66,2.4$ for $\alpha=2,3,4,5$, respectively. 
The scaling factors are different from $s= \alpha^2/8=0.5,1.125,2,3.125$, which would result from Eq.~\eqref{tstar} for $\alpha=2,3,4,5$. 
This difference can be attributed to finite-size effects. To see this, let us rewrite Eq.~\eqref{dv} in the variable $x = \ln v$.
Using Itô's lemma, we get 
\begin{equation}
    dx_j(t) = -(\beta + \sigma^2/2 - \beta e^{-x_j(t)}) dt + \sigma dB_j(t)
\end{equation}
or equivalently  
\begin{equation}
    dx_j(\tilde{t}) = -\frac{1}{2} \left(\alpha - (\alpha-1)e^{-x_j(t)}\right) d\tilde{t} +  dB_j(\tilde{t})
\end{equation}
where $\alpha = 1 + 2\beta/\sigma^2$ and $\tilde{t} = \sigma^2 t$.
We see that this equation is of the form~\eqref{drif}, which
describes Brownian motion in the presence of drift, discussed in section \ref{sec:example1}. In the present case, the drift is related to $\alpha$ as
\begin{equation} \label{alpha_eff}
 -2 \mu(x)= \alpha - (\alpha-1) e^{-x} \equiv \alpha_{\rm eff}(x).
\end{equation}
The value of drift approaches $-\alpha/2$ for $x\rightarrow \infty$; for finite $x$, however, it is equal to $-\alpha_{\rm eff}(x)/2$. For systems of finite size, the drift should be calculated
for the position of the leaders, which are located at finite $x$.
The position of the leader, $x_1$, in a population of $N$ elements can be estimated as a quantile of the stationary distribution for $1-1/N$ that gives 
\begin{equation}\label{better.approx}
    \frac{\gamma\left(\alpha,(\alpha-1) e^{-x_1}\right)}{\Gamma(\alpha)} = 1/N.
\end{equation}
where $\gamma(\alpha,y)=\int_0^y t^{\alpha-1} e^{-t} dt$ is the
lower incomplete Gamma function.
The expression on the left-hand side is the value of
the complementary cumulative distribution
function for $v_1=e^{x_1}$ of the inverse gamma distribution~\eqref{inv_gamma}, which is the stationary state.
From this equation, one can estimate that
\begin{equation}\label{approx}
    e^{-x_1} \approx \frac{\left(\Gamma(\alpha+1)/N\right)^{1/\alpha}}{\alpha-1}
\end{equation}
which, after inserting into Eq.~\eqref{alpha_eff}, gives
\begin{equation}
\alpha_{\rm eff}(x_1) \approx \alpha - \left(\Gamma(\alpha+1)/N\right)^{1/\alpha} .
\end{equation}
Setting $N=5000$, as in the simulation shown in Fig. \ref{fig:BM20},
we obtain from~\eqref{approx}
\begin{equation}
s_{\rm eff} = \frac{\alpha^2_{\rm eff}(x_1)}{8} \approx
0.490, 1.047, 1.745, 2.560
\end{equation}
and from a numerical solution of~\eqref{better.approx}
\begin{equation}
s_{\rm eff} = \frac{\alpha^2_{\rm eff}(x_1)}{8} \approx
0.490, 1.044, 1.731, 2.513
\label{seff}
\end{equation}
for $\alpha=2,3,4,5$, respectively. These numbers are extremely
close to those used as scaling factors in Fig.~\ref{fig:BM20},
which means that finite-size corrections explain the deviation
of the scaling factor from its asymptotic value. The remaining deviations (which are less than 2\%) can come from statistical errors and other approximations mentioned in our argumentation. For instance, the numerical fit is actually an average of the $n=1$ and $n=10$ fits, which should have slightly different effective $\alpha_{\rm eff} $ due to the variation of the positions of the individual largest particles.

To conclude this example, it is worth noting that in this model $x_j(t)-x_j(0)$ is the cumulative growth rate above the average inflation level for the entire system (see Eq.~\eqref{relative}).

\subsection{Example 4: Overlap ratio for positive stationary Kesten processes}\label{sec:example4}

Another example where the same universality of the rank statistics of the leading particles can be expected is a system of $N$ particles 
whose positions evolve according to Kesten processes~\cite{K}
\begin{equation} \label{kp}
    w_j(k) = a_j(k) w_j(k-1) + b_j(k).
\end{equation}
The random variables $a_j$ and $b_j$ are identical
independent processes consisting of independently identically distributed positive random numbers. If initially the particles lie in the positive half-axis, then they will stay there during the whole process.

\begin{figure}
    \centering
    \includegraphics[width=0.6\textwidth]
    {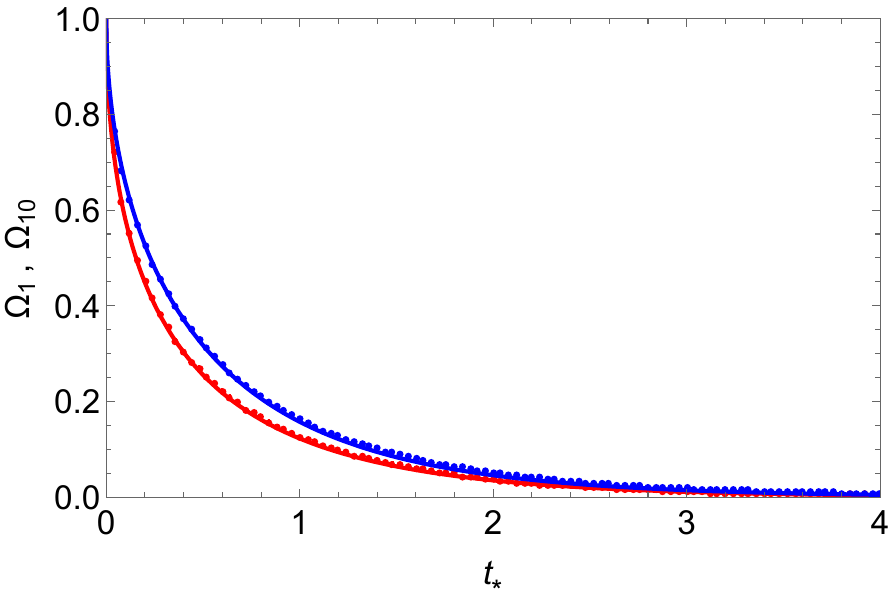}
    \caption{Overlap ratio $\langle \Omega_n(t) \rangle$
    for $n=1,10$ for the gas of $N=5000$ particles whose positions are
    generated by independently identically distributed Kesten processes $w(t) = a(t)w(t-1)+b(t)$ 
    with $a,b$ being  independently identically distributed log-normal processes $a\sim \ln N(\mu_a,\sigma_a^2)$
    with $\mu_a=-0.1$, $\sigma_a^2=0.1$, and $b\sim \ln N(\mu_b,\sigma_b^2)$  with $\mu_b=-5$, $\sigma_b^2=0.01$.
    The asymptotic properties of the stationary distribution has a Pareto tail with the exponent
    $\alpha=-2 \mu_a/\sigma_a^2 =2$. The overlap ratio is plotted 
    against the universal variable $t_* = \alpha^2 t/8$, where
    $t=k\sigma^2_a$. The data points are averaged over $10^4$ simulations. The numerically calculated overlap 
    ratios for $n=1$ (red dots) and $n=10$ (blue dots) are compared to the analytic curves $\Omega_1(t_*)$ in~\eqref{omega1} (red line, lower curve) and $\Omega_*(t_*)$ in~\eqref{omega_star} (blue line, upper curve) derived
    for the diffusion on $\mathbb{R}_+$.
     \label{fig:kesten_omega}    
     }
\end{figure} 

The system has a stationary state if the conditions  $\langle \ln a \rangle < 0$ 
and $\langle b \rangle < \infty$ are satisfied. The first condition prevents that the multiplicative evolution is not exponentially growing while the second one implies that the evolution is not too strongly diffusive if L\'evy flights are present. We skipped the index $j$, here, because
the processes are identical.  
The stationary state has a distribution with a Pareto tail $\tilde{p}_{\rm stat}(w) dw \sim w^{-1-\alpha} dw$ if $\langle a^\alpha \rangle = 1$, $\langle a^{\alpha+1} \rangle < \infty$ and $\langle b^\alpha \rangle < \infty$. Actually, there are some additional conditions to exclude interferences between $a$ and $b$, see~\cite{K,B} for details.

As an example, consider $a_j(k)$ and $b_j(k)$ identical independent lognormal processes $a_j \sim \ln \mathcal{N}(\mu_a, \sigma_a^2)$
and $b_j \sim \ln \mathcal{N}(\mu_b,\sigma_b^2)$, with $\mu_a<0$.
In this case, the associated Kesten process has a stationary distribution with a Pareto tail with index $\alpha = -2 \mu_a/\sigma_a^2$. Physical time $t$
can be calculated from the discrete time $k$ using the formula $t = k \Delta t = k \sigma_a^2$. 

The result of the numerical simulation is shown in Fig.~\ref{fig:kesten_omega}.
The overlap ratios are very well described by
the analytical formulas~\eqref{omega1} and~\eqref{omega_star}
that have been used for diffusion on $\mathbb{R}_+$. The result can be intuitively understood as follows. Iterating~\eqref{kp} leads to
\begin{equation}
   w(k) = a(k) \ldots a(1) w(0) + \sum_{s=1}^{k} a(k) \ldots a(s+1) b(s)  
\end{equation}
The associated process $x(k) = \ln w(k)$ is for large $x$ mainly 
shaped by the leading product of the random variables $a(j)$ which after taking the logarithm makes $x(k)$ to behave for large $x$
as an almost additive process with negative drift. For smaller $x$, on the other hand, a soft barrier is created by adding a positive value $b$ in each step of the process, which makes values of $x(k)$ bounce towards larger values.
In many respects, the situation is similar to diffusion, discussed in Sect.~\ref{sec:univ}, 
which has a constant negative drift
asymptotically for $x\rightarrow +\infty$ and a soft reflective barrier at some finite $x$.

\subsection{Example 5: Overlap ratio for Ornstein-Uhlenbeck 1D gas}\label{sec:example5}

All the examples discussed so far describe stochastic processes that asymptotically behave
for large $x$ as a diffusion with constant negative drift. In Example 1, 
the process is formulated directly in the variable $x$, while in Examples 3-4 in $w=e^x$. 
In all cases, changes in $\Delta x_j$'s behave as i.i.d. variables and lead to
a stationary distribution for $x$ that has an exponential tail or, equivalently, for $w$, a Pareto tail. 
We have seen that in all cases both the overlap ratios $\langle \Omega_1(t) \rangle $ and
$\langle \Omega_\infty(t) \rangle$ are very well described by the theoretical expressions derived for the diffusion
on $\mathbb{R}_+$; see Eqs.~(\ref{omega1}) and (\ref{omega.t}), respectively. 

\begin{figure}
    \centering
    \includegraphics[width=0.6\textwidth]
    {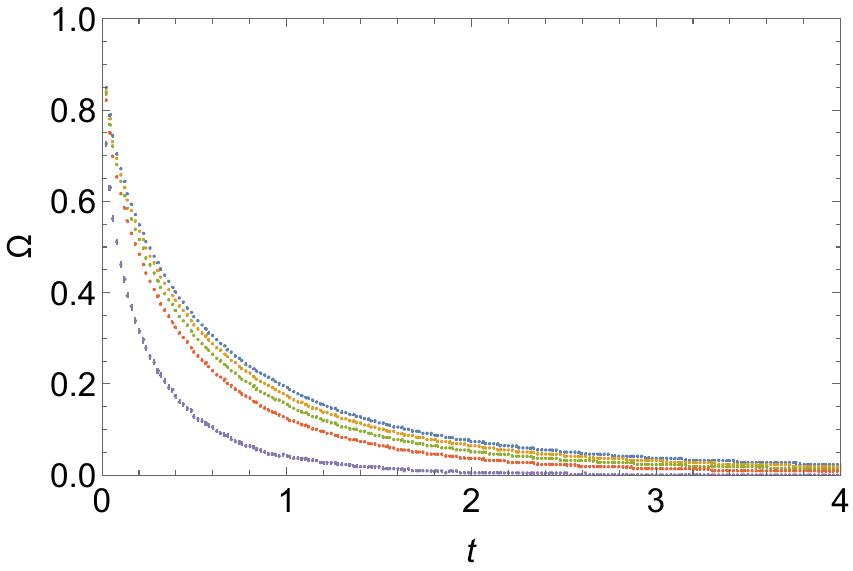}
    \caption{Overlap ratio for $n=1,25,50,75,100$ (from bottom to top)
    obtained in Monte-Carlo simulations of a gas of $N=10^4$ particles undergoing Ornstein-Uhlenbeck 
    process (\ref{OE}) with $\kappa=1$. Data points are obtained by averaging over $10^4$ simulations.
     \label{fig:ou_omega}    
     }
\end{figure} 

In the present example, we consider a gas of $N$
particles undergoing the Ornstein-Uhlenbeck stochastic evolution
\begin{equation} \label{OE}
d x_j(t) = -\frac{1}{2} \kappa x_j dt + dB_j(t)
\end{equation}
where $\kappa$ is a positive constant, and $B_j(t)$, $j=1,\ldots,N$ are independent Wiener processes, as before. 
The drift is linear in $x$: $\mu(x)=-\frac{1}{2}\kappa x$. The Fokker-Planck equation reads:
\begin{equation}
	\partial_t p(x,t)= \frac{1}{2} \kappa \partial_x \left(x p(x,t)\right) +
    \frac{1}{2} \partial_x^2 p(x,t) .
\end{equation}
The stationary distribution is given by a Gaussian distribution 
\begin{equation}
	p_{\rm stat}(x)= \sqrt{\frac{\kappa}{2\pi}} \exp\left[-\frac{\kappa x^2}{2}\right], 
\end{equation}
and $w=e^x$ is characterized by a log-normal distribution, which is also very common in economic reality, 
where it is often referred to as Gibrat's law.
We performed Monte-Carlo simulations of a gas consisting of $N$ particles undergoing 
independent Ornstein-Uhlenbeck processes based on a discretized version of (\ref{OE})
\begin{equation}
x_j(k) = a x_j(k-1) + b g_j(k)
\end{equation}
where $a=1 - \kappa \Delta t/2$, $b=\sqrt{1-a^2}$, and $g_j(k)$ are independently and identically normally distributed 
random numbers with zero mean and unit variance. The result is shown in Fig.~\ref{fig:ou_omega}. As can be
seen, the overlap ratio behaves differently than that for diffusion in constant drift, which applies to
all previously discussed examples. The overlap ratio $\Omega_1(t)$ for the Ornstein-Uhlenbeck
diffusion is much more convex than the corresponding curve (\ref{omega1}), which very accurately describes all
previous examples. It drops much faster in a short time, meaning that the
position of the leader is not as persistent as it was in the other models. 
The dependence of $\Omega_n(t)$ on $n$ is also different. For constant drift, already for $n=10$, it was difficult to distinguish the shape of the function $\Omega_n(t)$ from the limiting one $\Omega_\infty$ (\ref{omega_star}), 
while here we see that for $n=100$ the overlap ratio has not yet converged to its asymptotic
form. These effects can be attributed to much faster mixing of the order of extreme values
of particles undergoing diffusion in the quadratic potential than in the linear potential.

\begin{figure}
    \centering
    \includegraphics[width=0.6\textwidth]
    {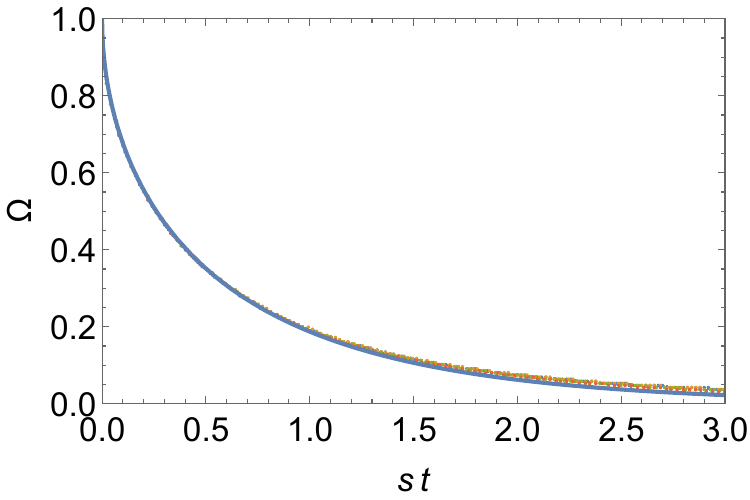}
    \caption{The same data as in Fig.~\ref{fig:ou_omega}, but presented differently.
The overlap factor $\Omega_n$ for $n=25,50,75,100$ is plotted as a function of $s_n t$, where $s_n$ are the $n$-dependent scale factors: $s_n = 1.4, 1.2, 1.1, 1.0$ for $n=25,50,75,100$. With this choice of scale factors, the data points for different $n$ almost collapse into a single curve. The plot also shows the function ${\rm erfc}(\sqrt{t/1.15})$ as a solid line.
     \label{fig:ou_collapse_erfc}    
     }
\end{figure}

We show in Fig.~\ref{fig:ou_collapse_erfc} the same data as in Fig.~\ref{fig:ou_omega}, 
for $n=25,50,75,100$, but plotted against $s_n t$ with scale factors dependent on $n$. We see that
the scaling makes the data for different $n$ collapse into a single shape. For small $t$, this
shape is consistent with ${\rm erfc}( a\sqrt{t})$. For large $t$, there are apparent deviations,
but they can be attributed to finite size effects, which are significant, as $n/N$ is on the order of a percent in this case
(compare Fig.~\ref{fig:omega_1_finite_N}). To summarize, it is likely that the asymptotic form of the overlap ratio for the diffusion of $N$ particles
in a quadratic potential for $1 \ll n \ll N$ will also be given by a complementary error function; however,
 the convergence to the asymptotic form is much slower than in a linear potential, somewhat like
the convergence to the Gumbel distribution of extreme value statistics, which is much slower
for a Gaussian distribution than for an exponential distribution. 
The collapse of data observed in Fig.~\ref{fig:ou_collapse_erfc} raises the question of whether
the $n$-dependence in the rescaling $t\to s_n t$ persists or will eventually vanish. 
However, we will be able to answer these questions unambiguously only after performing analytical
calculations for the Ornstein–Uhlenbeck process.

\subsection{Example 6: Overlap ratio in the presence of asymptotically vanishing drift \label{e6}}

\begin{figure}
    \centering
    \includegraphics[width=0.6\textwidth]{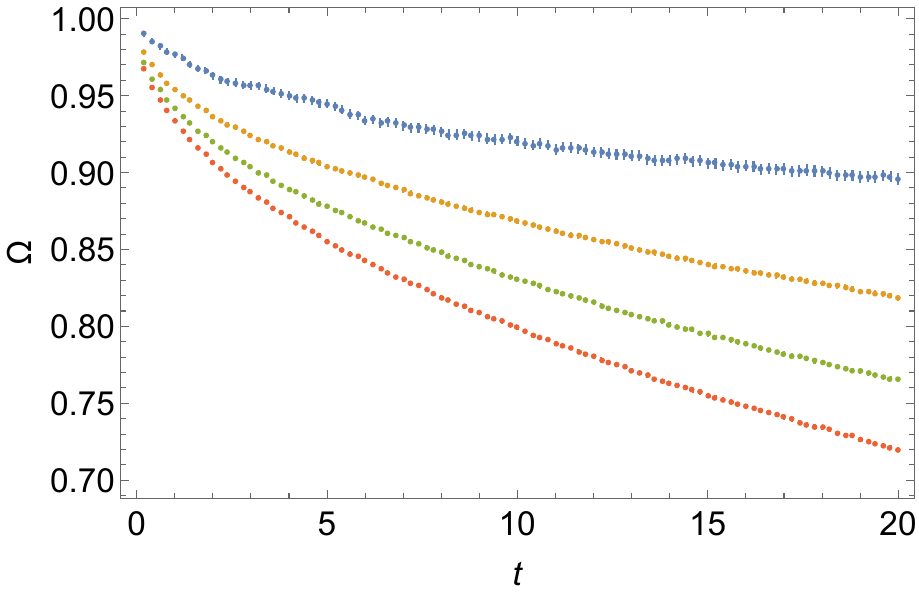}
    \caption{Overlap ratio for $n=1,10,20,30$ (from top to bottom)
    obtained in Monte-Carlo simulations of a gas of $N=10^4$ particles undergoing diffusion 
    on the interval $[1,\infty)$ with a drift $\mu(x) = -(\gamma+1)/x$, with $\gamma=2$, 
    and reflective wall at $x=1$. Data points are obtained by averaging over $10^4$ simulations.
     \label{fig:heavy_gamma_2}    
     }
\end{figure}

As a final example, we consider diffusion in a logarithmic potential $V(x) \sim \ln x$ for $x \rightarrow \infty$,
which generates a much weaker drift than in the examples discussed so far. Clearly, one can expect the order statistics to be more persistent in this case.
More precisely, we will consider diffusion in a potential $V(x) = (\gamma + 1) \ln x$ 
bounded from below by a reflective wall at $x=1$.
In this case, the drift tends to zero like $\mu(x) =-V'(x) = -(\gamma+1)/x$
as $x$ increases. For $\gamma>0$, the system has a stationary state with 
an extremely extended probability density
\begin{equation}{\label{xg}}
    p_{\rm stat}(x) dx = \Theta(x-1) \frac{\gamma dx}{x^{\gamma+1}}. 
\end{equation}
The results of Monte Carlo calculations of the overlap ratio for a system with $\gamma=2$ are shown
in Fig. \ref{fig:heavy_gamma_2}.

The most striking difference compared to the previous cases is that $\Omega_1(t)$ decreases as a function of $t$
slower than $\Omega_n(t)$ for $n>1$. The smaller $n$, the slower $\Omega_n(t)$ decreases with time.
This result can be easily understood. In the presence of the power-law tail (\ref{xg}) the positions
of the leaders are very persistent. The leaders are located near $x_{\rm top} \sim N^{1/\gamma}$ 
and separated from the bulk by long gaps. It takes a long time for the diffusion to cover the distance
between the leaders. Once a particle becomes a leader, it stays a leader for a long time. 

This model is completely unrealistic from the point of view of growth rate dynamics. 
We included it to illustrate the limits of the universality class as well as the obvious relation between the persistence of
leader-order statistics and the asymptotic properties of the drift.

\section{Conclusions}\label{sec:conclusions}

We have proposed an analytical method to calculate the average overlap coefficient between
top-$n$ lists in diffusion models or rank dynamics. The method is based on the
relationship between the probability of reshuffling and the average overlap ratio. 
We used this relation to derive analytical expressions for the average overlap ratio $\Omega_n(t)$
(\ref{omega_star_n}) and its limiting form $\Omega_*(t)$ for $n\rightarrow \infty$ (\ref{omega_star}) 
for diffusion with constant drift on the half axis in the presence of a reflective wall. The limiting
form of the overlap ratio has an extremely simple form $\Omega_*(t)={\rm erfc}(a\sqrt{t})$ with a single
free parameter $a$. When applying this formula to empirical data, the value of the parameter $a$ 
can be fixed by the procedure based on~\eqref{emp.fit}. This parameter is related to
drift $\mu<0$ and the diffusion coefficient $\sigma^2/2$: $a=|\mu|/\sigma^2$.
In the limit of vanishing drift $\mu\to0$, the reshuffling of the top ranks disappears $\Omega_*(t)\to1$. 
In this limit, the stochastic fluctuations are too weak to mix the top $n$ with the remaining participants; 
in particular, the mixing time scale grows to infinity. A similar effect is observed in Example 6, 
where we considered diffusion with asymptotically vanishing drift; see \ref{e6}. 

We have also observed that $\Omega_n(t)$ 
for $n\ge 10$ is practically indistinguishable from $\Omega_*(t)$; see Fig. \ref{fig:omega_th_vs_mc}. 
We conjecture that these results are universal for a broad class of models where the dynamics of leaders can be viewed as independent or weakly correlated random walks in the presence of a constant negative drift. To corroborate this conjecture, we studied different models for which we measured $\Omega_n$ and compared them to the analytical curves obtained for the prototype model with a constant drift on the positive half-axis.
The numerical results in all models studied lie on top of the analytically predicted curves;
see Figs.~\ref{fig:Omega_mux},\ref{fig:BM20},\ref{fig:kesten_omega}. The order-statistics dynamics observed in these
models is characteristic of diffusion, with the stationary distribution having an exponential tail,
which is related to an asymptotically constant drift. 

We observed a universality of the analytical results by comparing them with empirical data and numerical simulations. In particular, the reason for this universality can be understood on theoretical grounds. The asymptotic results for $N\rightarrow \infty$ are shaped by the behavior of the heat kernel $W$, see~(\ref{Womega}),
in the neighborhood of extreme values (\ref{xi_zeta}), where the ranking leaders are located. 
The extreme values scale with $\ln N$ and depart from the
finite part of the distribution. The results concerning top-order statistics are entirely dependent on the asymptotic properties of the tail of the distribution, which decays exponentially
for large $x$ in the case of the models considered.

Using diffusive models as models
of growth rate dynamics, we can apply them to explain the dynamics of order statistics for a quantity $w=e^x$, 
for which $x$ is the growth rate, because the map $x \rightarrow e^x$ is monotonic and preserves the order.
When the probability distribution of $x$ has an exponential tail, the probability distribution of $w$ has a Pareto tail.
This makes us believe that the formula for the overlap ratio $\Omega_*(t)$ in~\eqref{omega_star} 
should apply to many economic problems, such as wealth distribution or company size distribution, 
where Pareto tails are observed. In economic reality, one cannot expect stationarity, so one
should apply this approach carefully. For example, there is an obvious inflation factor in the
wealth distribution, which makes it difficult to directly compare the wealth distribution today with that of, say, ten years ago. 
But what one can do is first express individual wealth in units given by the current average
population wealth to make the inflation factor drop, as is done in the Bouchaud-Mézard model \cite{BM}. 
What remains is then the growth rate dynamics relative to the inflation rate. This residual growth rate can be considered quasi-stationary 
when limited to a time interval of an appropriate length. This time interval must be short enough to assume quasi-stationarity 
and long enough for the system to reach momentary equilibrium. Whether such an interval exists is a matter of the relative rate of
endogenous equilibrating processes versus exogenous changes.  
The assumption of quasi-stationarity seems plausible when the equilibration processes are faster than the others.

It can be argued that the wealth growth rates of the richest are independent of each other.
The leaders in the wealth ranking are those who show initiative and take independent actions, thanks to which their wealth grows faster than that of others. You cannot be a leader if you imitate. In other words, the growth rates of leaders are independent. 
The reason for the appearance of a negative drift in the growth rates of leaders can also be understood. It
comes from the conservation of total wealth (which, in terms of population wealth per capita, is constant), which means that
the wealth of ranking leaders cannot grow indefinitely.
Similarly, one can argue that the growth rates of leaders in other economic rankings are independent of each other. 
In summary, if we apply Gibrat's rule of proportionate growth \cite{G} to describe the dynamics of the growth rate relative
to the common inflation rate, we obtain the diffusion model we have discussed here. This is why we expect to observe the universal
behavior of the overlap ratio and the reshuffling of top ranks, characteristic of the diffusion of growth with a negative drift
in many economic rankings. 

A classic result in the field of extreme value statistics is the classification of probability distributions with
respect to max-stability. According to the Fisher-Tippet theorem \cite{FT}, there are three classes
that produce different order statistics of leaders. 
As far as the dynamical aspects of order statistics are concerned, distributions from the same class lead to
different order statistics dynamics, as we have seen in this work. Indeed, diffusion in the presence of constant drift
produces an exponential stationary distribution, and Ornstein-Uhlenbeck diffusion produces a normal distribution, 
both belonging to the Gumbel max-stability class. As we have seen, however, the overlap ratio that captures the order statistics of
the top $n$ ranks behaves differently in the two cases. In particular, $\Omega_1(t)$, {\em i.e.} , the probability that the leader
will be the leader after time $t$, is identical in all four examples
described in Sect.\ref{sec:example1}-\ref{sec:example4},
as in the case of diffusion with constant negative drift, Eq. \eqref{omega1}, see Figs. \ref{fig:Omega_mux},\ref{fig:BM20},\ref{fig:kesten_omega}, while it is different for the Ornstein-Uhlenbeck
diffusion; see Sect. \ref{sec:example5} and Fig. \ref{fig:ou_omega}. 
Thus, within the same max-stability class, there are different classes of order
 statistics dynamics. So, within the same max-stability class,
there are different classes of order statistics dynamics. Of course, the classification of these
classes will also depend on the diffusion type; it will certainly be different for fractional diffusion
or Lévy flights than for Brownian motion. 

To our knowledge, the dynamic aspects of order statistics have not yet been systematically investigated, so
there are many open questions. 
The challenges include: to prove the universality that we conjectured here for the class of 1D diffusion models with an asymptotically constant drift; 
a related problem is to determine the asymptotic convergence to the limiting expressions for
the overlap ratio as $N$ goes to infinity and then $n$ goes to infinity;
to calculate analytically the overlap ratio for
the Ornstein-Uhlenbeck model; to
classify universality classes according to the asymptotic properties of the potential in which diffusion occurs and according to the type of diffusion; to perform empirical research on top-$n$ order statistics dynamics, 
by measuring the overlap ratio for rankings in the real world. 

Another category of stochastic processes that shape complex systems in many situations is those based on growth rather than reshuffling.
In many systems, growth is driven by preferential attachment \cite{Y,S,BA} leading to heavy-tailed statistics. The values of local quantities
describing the state of the system (for instance, degree of nodes in network models) generated in such growth processes are correlated with
attachment time \cite{KR}, and in this case the reshuffling of extreme values is very limited, which is reflected in rank statistics \cite{BGFSBBB,GB}.
It would be interesting to calculate the overlap ratio for such growing processes, as was done here for random reshuffling. 
The overlap ratio could then serve as a benchmark that would help to distinguish the influence of stochastic reshuffling
versus growth on the top-rank statistics. 

\begin{acknowledgements}
\label{ack}

We thank Jean-Philippe Bouchaud for discussions and valuable suggestions, and Krzysztof Malarz for creating the figure \ref{fig:empirical} based
on data from \cite{BKMS}. MK is partially funded by the Australian Research Council through
the Discovery Project grant DP250102552, by the Alexander-von-Humboldt Foundation, and by the Swedish Research Council under grant no. 2021-06594 while the author was in residence at the Mittag-Leffler Institute in Djursholm, Sweden, during November/December of 2024. MK is also grateful to the Faculty of
Physics and Applied Computer Science at the AGH University of Krakow and the Faculty of Physics of the University of Duisburg-Essen for their hospitality during his sabbatical.
This research was partially supported by
a subsidy from the Polish Ministry of Science and Higher 
Education.

\end{acknowledgements}

\appendix

\section{Derivation of the heat kernel} \label{app:HK_derivation}

In the first step to solve~\eqref{PDE} and~\eqref{boundary}, we write the probability density $p(x,t) $ in terms of a mixed Fourier-Laplace transform
\begin{equation}\label{LFA1}
p(x,t) = \int d \omega dk \exp[- \omega t +i k x] \hat{p}(k,\omega).
\end{equation}
When plugging this into the Fokker-Planck equation~\eqref{PDE} we arrive at
\begin{equation} \label{LFA2}
0= \int d \omega dk \left(\frac{1}{2} k^2 - \frac{i\alpha}{2} k - \omega \right) \hat{p}(k,\omega) = \int d \omega dk \frac{1}{2} 
\left(k - k_+(\omega)\right)\left( k - k_-(\omega) \right) \hat{p}(k,\omega),
\end{equation}
where 
\begin{equation} \label{kpm}
k_\pm(\omega) = \frac{1}{2}\left( i \alpha  \pm  \sqrt{8\omega - \alpha^2}\right) .
\end{equation}
Therefore, the Fourier-Laplace transform $\hat{p}(k,\omega)$ must have the form
\begin{equation}
    \hat{p}(k,\omega) = \delta(k-k_+(\omega)) 
    \hat{p}_+(\omega) + \delta(k-k_-(\omega)) \hat{p}_-(\omega),
\end{equation}
where $p_\pm(\omega)$ are still some unknown functions of $\omega$. 

Performing the integral over $k$ in~\eqref{LFA1} yields
\begin{equation}
p(x,t) = \int_{\alpha^2/8}^\infty d \omega  
e^{- \omega t}\left(  e^{ik_+(\omega)x} \hat{p}_+(\omega) +
e^{ik_-(\omega)x} \hat{p}_-(\omega) \right) .
\end{equation}
The lower terminal of $\omega$ at $\alpha^2/8$ takes into account the fact that  the factors $e^{i k_\pm(\omega) x}$ 
have an oscillating (trigonometric) part and a common exponential decay due to the constant drift, 
or equivalently that~\eqref{kpm} has the form
\begin{equation} 
i k_\pm(\omega) = \frac{1}{2}\left(-\alpha \pm i \kappa\right),
\end{equation}
where $\kappa = \sqrt{8\omega - \alpha^2}$ is a non-negative real number. After substituting $\omega=(\kappa^2+\alpha^2)/8$, we arrive at
\begin{equation} \label{pkappa}
p(x,t) = \int_{0}^\infty d \kappa   
\exp\left(- \frac{(\kappa^2+\alpha^2) t}{8}\right) 
\left[ \exp\left(\frac{(-\alpha+i\kappa) x}2\right) \hat{q}_+(\kappa) +
 \exp\left(\frac{(-\alpha-i\kappa) x}2\right) \hat{q}_-(\kappa) \right] ,
\end{equation}
where $\hat q_\pm(\kappa)$ are derived from $p_\pm(\omega)$ by the change of variables.

Next, we try to satisfy the boundary condition~\eqref{boundary}, which becomes 
\begin{equation}
    0 = \alpha p(0,t) + \partial_x p(0,t) = \int_0^\infty d\kappa \exp\left(- \frac{(\kappa^2+\alpha^2) t}{8}\right)  \left[ \frac{\alpha + i \kappa}{2} \hat{q}_+(\kappa) + 
    \frac{\alpha - i \kappa}{2} \hat{q}_-(\kappa) \right].
\end{equation}
Therefore, the functions $\hat q_+(\kappa)$ and $\hat q_-(\kappa)$ must be related. In fact,
the expression in the square brackets must be identically equal to zero because the integral must vanish for all $t>0$. We can express the two functions $\hat q_\pm(\kappa)$ in terms of a single but otherwise unknown function $\hat{q}(k)$,
\begin{equation}
   \hat{q}_\pm(\kappa) = \frac{ \mp \alpha + i \kappa}{2} \frac{\hat{q}(\kappa)}{2i}
\end{equation}
Then, the probability density~\eqref{pkappa} is equal to
\begin{equation} 
p(x,t) = \int_{0}^\infty d \kappa   
\exp\left(- \frac{(\kappa^2+\alpha^2) t}{8}\right) 
\left[ \frac{-\alpha + i \kappa}{2} \exp\left(\frac{(-\alpha+i\kappa) x}2\right)  
+ \frac{\alpha + i \kappa}{2} 
 \exp\left(\frac{(-\alpha-i\kappa) x}2\right)  \right] \frac{\hat{q}(\kappa)}{2i}
\end{equation}
or equivalently
\begin{equation} \label{pq}
p(x,t) = \partial_x \int_{0}^\infty d \kappa   
\exp\left(- \frac{(\kappa^2+\alpha^2) t}{8} - \frac{\alpha x}{2} \right) \sin\left(\frac{\kappa x}{2}\right)
 \hat{q}(\kappa)  .
\end{equation}

Finally, we need to implement the initial condition that fixes the function $\hat{q}(k)$. Putting $t=0$ in the last equation, we have
\begin{equation} 
p_0(x)=p(x,0) = \partial_x  \int_{0}^\infty d \kappa \exp\left(-\frac{\alpha x}{2}\right)  
\sin\left(\frac{\kappa x}{2}\right)
 \hat{q}(\kappa)  .
\end{equation}
Integrating both sides with respect to $x$ gives
\begin{equation} \label{x}
\int_0^x p(\xi,0) d\xi =  \exp\left(-\frac{\alpha x}{2}\right) \int_{0}^\infty d \kappa  
\sin\left(\frac{\kappa x}{2}\right)
 \hat{q}(\kappa)  .
\end{equation}
The left hand side is equal to the cumulative distribution function
\begin{equation}
P_-(x,t) = \int_0^x d\xi p(\xi,t)
\end{equation}
for $t=0$. The cumulative distribution function $P_-(x,t)$ is the probability that the particle in time $t$
is in the interval $[0,x]$. Multiplying both sides of \eqref{x} by 
$\exp\left(\alpha x/2\right)$ and introducing a function
\begin{equation} 
q(x) \equiv 
\exp \left( \frac{\alpha x}{2} \right) \int_0^x d\xi p(\xi,0) =
\exp \left( \frac{\alpha x}{2} \right) P_-(x,0)
\end{equation}
we obtain
\begin{equation} 
q(x) = \int_0^\infty d\kappa \sin\left(\frac{\kappa x}{2}\right) \hat{q}(\kappa) .
\end{equation} 
The last equation means that $q(x)$ is just the Fourier sine-transform of $\hat{q}(\kappa)$.
We can use the inverse sine-transform to determine $\hat{q}(\kappa)$ from
$q(x)$:
\begin{equation} 
\hat{q}(\kappa) = \frac{1}{\pi} \int_0^\infty dx \sin\left(\frac{\kappa x}{2}\right) q(x) = 
\frac{1}{\pi} \int_0^\infty dx \sin\left(\frac{\kappa x}{2}\right) \exp \left( \frac{\alpha x}{2} \right) P_-(x,0) .
\end{equation} 
Inserting the expression on the right hand side for $\hat{q}(\kappa)$ to~\eqref{pq} and integrating both sides over $x$ we find an equation for $P_-(x,t)$ 
\begin{equation}
\begin{split}
P_-(x,t) =& \int_{0}^\infty d \kappa   
\exp\left(- \frac{(\kappa^2+\alpha^2) t}{8} - \frac{\alpha x}{2} \right) \sin\left(\frac{\kappa x}{2}\right)
\frac{1}{\pi} \int_0^\infty dy \sin\left(\frac{\kappa y}{2}\right) \exp \left( \frac{\alpha y}{2} \right) P_-(y,0) \\
=&\exp\left(- \frac{\alpha^2 t}{8}\right)
\int_{0}^\infty d y \exp \left( \frac{\alpha (y-x)}{2} \right) P_-(y,0)  
\frac{1}{\pi} \int_0^\infty d\kappa
\exp\left(- \frac{\kappa^2 t}{8}  \right) \sin\left(\frac{\kappa x}{2}\right)
 \sin\left(\frac{\kappa y}{2}\right) .
\end{split}
\end{equation}
In the second line, we interchanged the order of integration. The integral
over $\kappa$ can be done, as it is a sum of two Gaussian integrals. It gives
\begin{equation}
\frac{1}{\pi} \int_0^\infty d\kappa
\exp\left(- \frac{\kappa^2 t}{8}  \right) \sin\left(\frac{\kappa x}{2}\right)
 \sin\left(\frac{\kappa y}{2}\right) = \frac{1}{\sqrt{2\pi t}}
 \left[ \exp\left(-\frac{(x-y)^2}{2t}\right) - 
 \exp\left(-\frac{(x+y)^2}{2t}\right)  \right].
\end{equation}
Substituting this into the above equation, we obtain the equation for the evolution of the cumulative density function
\begin{equation} \label{PP}
P_-(x,t) = \int_{0}^\infty d y P_-(y,0) L(y,x,t) 
\end{equation}
with 
\begin{equation}
L(y,x,t) = \frac{1}{\sqrt{2\pi t}} \exp\left(- \frac{\alpha^2 t}{8}\right)
\exp \left( \frac{\alpha (y-x)}{2} \right) 
\left[ \exp\left(-\frac{(x-y)^2}{2t}\right) - 
 \exp\left(-\frac{(x+y)^2}{2t}\right)  \right]  . 
\end{equation}
The function $L(y,x,t)$ is related to the function $F(y,x,t)$ in~\eqref{F} as follows
\begin{equation}
L(y,x,t) = -\partial_y F(y,x,t) = \frac{1}{\sqrt{2\pi t}} 
\exp\left(-\frac{(2(y-x) - \alpha t)^2}{8t}\right) - 
\frac{e^{-\alpha x}}{\sqrt{2\pi t}}   \exp\left( - \frac{(2(y+x) - \alpha t )^2}{8t}\right)  . 
\end{equation}

We plug this relation into~\eqref{PP} and integrate by parts to find
\begin{equation} 
P_-(x,t) = \int_{0}^\infty d y p(y,0) F(y,x,t) .
\end{equation}
The boundary terms vanish as either $P_-(y=0,0)=0$ or $F(y=\infty,x,t)=0$.
Taking the derivative of both sides with respect to $x$ and using~\eqref{FK},
we finally arrive at \eqref{green}.

\section{Sum rules} \label{app:sum_rules}

We first check the sum rules~\eqref{identities}. When exploiting the expression in terms of the contour integral in $z$ in~\eqref{trans.prob.fixed.y}, we can understand the sum combined with the derivative in $x$ as the explicit form of the product rule. For the second expression in~\eqref{trans.prob.fixed.y}, this is
\begin{equation}
\sum_{j=1}^N P(k;j;t|y_1,\ldots,y_N) = 
 \oint  \frac{dz}{2\pi i z^{k}(1-z)} \int_0^\infty dx
 \partial_x \left(
\prod_{l=1}^N\big(z+(1-z) F(y_l,x,t)\big)\right) = 
\oint \frac{dz (1-z^N)}{2\pi i z^{k}(1-z)} = 1.
\end{equation}
In the second equality, we used $F(y,0,t)=0$ and $F(y,\infty,t)=1$, and in the last one we exploited the geometric sum in combination with $k\leq N$. 

The second identity in~\eqref{identities} is also straightforward. The sum over $k$ is a geometric sum which gives
\begin{equation}
\begin{split}
\sum_{k=1}^N P(k;j;t|y_1,\ldots,y_N) & = 
\int_0^\infty dx  \oint \frac{dz}{2\pi i z^{N}} \frac{1-z^N}{1-z}
\partial_x F(y_j,x,t) 
\prod_{l\neq j}\big(z+(1-z) F(y_l,x,t)\big).
\end{split}
\end{equation}
The integrand is a meromorphic function in $z$ with the only poles in $z=0$ and $z=\infty$. When choosing the contour as a circle with $|z|>1$, then we notice that the integral over the first term proportional to $1/(1-z)$ must vanish as it is holomorphic at $z=\infty$ and the only poles lie inside the contour, meaning we can shift the contour to infinity where the integrand vanishes.

For the second term proportional to $z^N/(1-z)$ the pole at the origin is removed and the newly created pole at $z=1$ is simple for which the residue theorem yields
\begin{equation}
\begin{split}
\sum_{k=1}^N P(k;j;t|y_1,\ldots,y_N) = 
\int_0^\infty dx  
\partial_x F(y_j,x,t) 
\prod_{l\neq j}\big(1+0\cdot F(y_l,x,t)\big)= \int_0^\infty dx \partial_x F(y_j,x,t) =1
\end{split}
\end{equation}
The last equality is clear due to $F(y,0,t)=0$ and $F(y,\infty,t)=1$.

Now, we move on to prove~\eqref{identities_pi0}. We first apply the geometric sum
\begin{equation}
\label{SumRule3}
\begin{split}
    \sum_{k=1}^N P(k;j;t|  p_0 ) 
 = N \int_0^\infty dx \int_0^\infty dr W(r,x,t)p_0(r) \oint \frac{dz}{2\pi i z^j} \oint \frac{dw}{2\pi i w^N}\frac{1-w^N}{1-w} 
\big( w P_z(r) + (1-w) G_z(r,x,t) \big)^{N-1} .
\end{split}
\end{equation}
We recall the definitions~\eqref{PGs} and~\eqref{Gs}.
Like before, the integrand is meromorphic in $w$ with the only poles at the origin and infinity, allowing us to choose the contour along $|w|=2$, for example. However, when the integrand is divided into one term proportional to $1/(1-w)$ and another term proportional to $w^N/(1-w)$, the first term must vanish, as all singularities of this term lie inside the contour, while the second term can be evaluated at the only residue at $w=1$. Thence, it is
\begin{equation}
\label{SumRule4}
\begin{split}
    \sum_{k=1}^N P(k;j;t|  p_0 ) 
 =& N \int_0^\infty dx \int_0^\infty dr W(r,x,t)p_0(r) \oint \frac{dz}{2\pi i z^j} P_z^{N-1}(r)\\
& = N \int_0^\infty dx \int_0^\infty dr W(r,x,t)p_0(r) \binom{N-1}{j-1} P_0^{N-j}(r)(1-P_0(r))^{j-1}  .
\end{split}
\end{equation}
In the second line, we have carried the second contour integral over $z$. Finally, we substitute $R=P_0(r)\in[0,1]$ and find the desired result
\begin{equation}
\label{SumRule5}
\begin{split}
    \sum_{k=1}^N P(k;j;t|  p_0 ) 
 = N \int_0^\infty dx  W(r,x,t) \binom{N-1}{j-1} \overbrace{\int_0^1 dR R^{N-j}(1-R)^{j-1} }^{=(N-j)!(j-1)!/N!} = \int_0^\infty dx W(r,x,t)= 1.
\end{split}
\end{equation}

Finally, we consider the sum over $j$ in~\eqref{identities_pi0}, which can be calculated in a similar way. Firstly, we rewrite the function
in the parentheses of~\eqref{Ppi0} as follows
\begin{equation}
    w P_z(r) + (1-w) G_z(r,x,t) = z \left[w + (1-w) G(x,t)\right] + (1-z)
    \left[wP_-(r) + (1-w) G_-(r,x,t)\right]  
\end{equation}
where 
\begin{equation}
    G(x,t) = G_-(r,x,t)+G_+(r,x,t) = \int_0^\infty F(r',x,t) p_0(r') dr' .
\end{equation}
We note that $G(0,t)=0$ and $G(\infty,t)=1$, and also that
\begin{equation}
    dG(x,t) = dx \int_0^\infty W(r,x,t) p_0(r) dr .
\end{equation}
Using this observation, we evaluate the sum along the very same lines as in Eqs.~(\ref{SumRule3}-\ref{SumRule5}) and find
\begin{equation}
\begin{split}
    \sum_{j=1}^N P(k;j;t|  p_0 ) 
& = N \int_0^\infty dx \int_0^\infty dr W(r,x,t)p_0(r)\\
& \times \oint \frac{dw}{2\pi i w^k} \oint \frac{dz}{2\pi i z^N} \frac{1-z^N}{1-z}  
\big( z \left[w + (1-w) G(x,t)\right] + (1-z)
    \left[wP_-(r) + (1-w) G_-(r,x,t)\right]  \big)^{N-1} \\
& = N \int_0^\infty dx \int_0^\infty dr W(r,x,t)p_0(r)\oint \frac{dw}{2\pi i w^k} 
\left(w + (1-w) G(x,t)\right)^{N-1} = 1 ,
\end{split}
\end{equation}
which was to be shown.

\section{Asymptotic behavior of the integral \texorpdfstring{~\eqref{doublepole}}{} \label{app:omega_large_n}}

Let us rewrite~\eqref{doublepole} as
\begin{equation} \label{2p}
 \Omega_{*n}(t_*) = \frac{e^{-t_*}}{n\sqrt{\pi}} \int_{-\infty}^{\infty} ds e^{-s^2}
  \oint \frac{dz}{2 \pi i z}
\frac{q}{(1+z^{-1})(1+z)}
 \left(1 - (1+z^{-1})\sigma_-\right)^n 
\left[\left(1- (1+z)\sigma_+\right)^n-1\right] 
\end{equation}
where $q=q(s,t_*)$ (\ref{denom}) and
\begin{equation}
\sigma_\pm = \sigma_{\pm}(s,t_*) = \frac{e^{\pm 2s\sqrt{t_*}}}{q(s,t_*)} .    
\end{equation}
In the following discussion, we will take advantage of the inequality
\begin{equation}
     0<\sigma_{\pm}  <1 .
\end{equation}
The inequality on the left side is obvious, while the one on the right side can be proven using the integral representation 
of $q(s,t_*)$~\eqref{denom}
\begin{equation}
    q(s,t_*) = e^{\pm  2\sqrt{t_*}s} + \frac{2}{\sqrt{\pi}} \int_0^\infty
    e^{-(\lambda \pm s)^2 - t_*} \sinh(2\sqrt{t_*} \lambda) d\lambda > e^{\pm  2\sqrt{t_*}s}.
\end{equation}
Dividing both sides by $q(s,t_*)$, we get $\sigma_{\pm}<1$.

\begin{figure}
    \centering
    \includegraphics[width=0.6\textwidth]{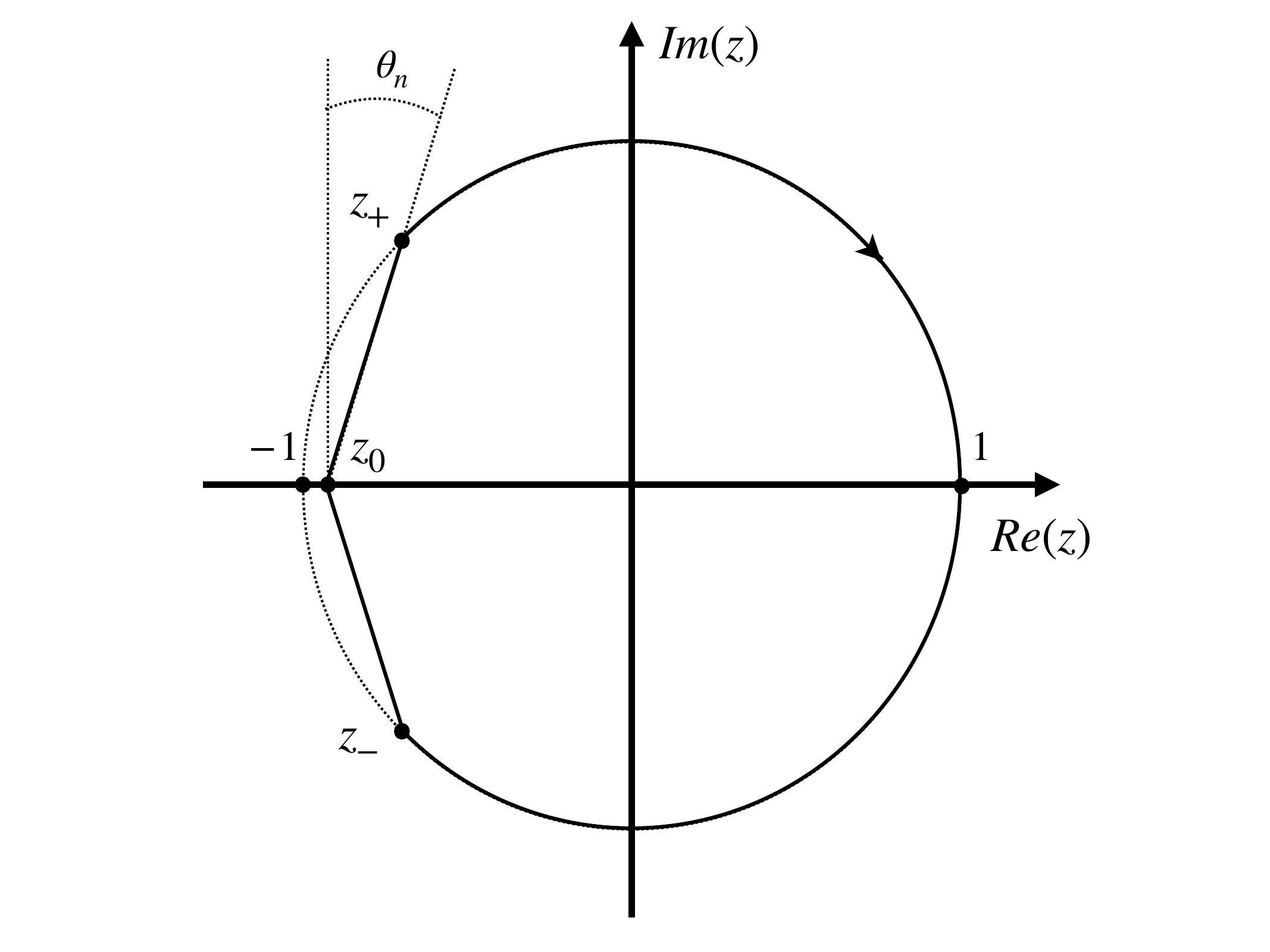} 
    \caption{The integration contour consists of two linear segments $[z_-,z_0]$, $[z_0,z_+]$
    and the unit circle arc going clockwise from $z_+$ to $z_-$. The angle $\theta_n$
    is between the vertical line (parallel to the imaginary axis) and the linear segment $[z_0,z_+]$. 
    For $n\rightarrow \infty$, it is $\theta_n\rightarrow 0$.\label{fig:icontour}}
\end{figure}

The original contour of integration over $z$ in~\eqref{2p} is a circle centered at zero with radius $\epsilon\ll 1$. The radius can be increased without affecting the integral, provided that no new singularities appear inside the increased contour. The radius can be increased almost to unity, but not to unity, because then the contour would pass through the double pole at $z=-1$. 
The idea is to deform the contour into a shape that is almost everywhere identical to the unit circle, except in the vicinity of the double pole, where it passes through the point at $z_0=-1 + 1/n$, which is slightly away from the pole; see the sketch in Fig. \ref{fig:icontour}. 
The points where the contour departs from the unit circle are
\begin{equation}\label{zpm}
z_{\pm}=-\left(1-\frac{1}{n^{2/3}}\right) \pm i \sqrt{1-\left(1-\frac{1}{n^{2/3}}\right)^2}.
\end{equation}
The two line segments connect these two points with the point $z_0=-1+1/n$. In this way, we avoid the singularity at $z=-1$. For $n\rightarrow \infty$, the integral over the arc of the unit circle tends to zero, as the leading contributions of it will be found at $z_\pm$, where the modulus of the integrand is of order $\mathcal{O}(n^{4/3} \exp[-n^{1/3}\sigma_-])$.

This means that the integrand exponentially tends to zero for all points on the arc of the unit
circle in the limit $n\rightarrow \infty$. We needed to choose the points~\eqref{zpm} 
with $1/n^{2/3}$ instead of the more natural $1/n$ scaling so that the contribution of the integration over the unit circle segment is suppressed. For the naive choice, this would not have been the case, as one is too close to the contributing point $z_0$.

What remains is the integral over the two line segments from $z_+$ to $z_0$ and from $z_0$ to $z_-$.
This integral is parameterized as follows:
\begin{equation}
z= -1 +\frac{1+ i e^{-i\,{\rm sign}(y)\theta_n} y}{n}\qquad {\rm with}\quad y\in[-n|z_+-z_0|,n|z_+-z_0|]
\end{equation}
where $\theta_n$ is the angle of the line segments to the imaginary axis. This angle is of the order $\mathcal{O}(n^{-1/3})$ and thus vanishes in the limit $n\to\infty$. The bounds of $y$ behave like $n^{2/3}$ and hence tend to infinity, so we effectively integrate at the end over the whole
line $z=-1+(1+ i y)/n$, with $y\in (-\infty,+\infty)$. The powers $(\ldots )^n$ in~\eqref{2p} approach exponential functions for $n\rightarrow \infty$, so we arrive
at~\eqref{exp2p}.

\end{document}